\documentclass[a4paper,fleqn]{cas-dc}

\usepackage[numbers,sort&compress]{natbib}
\usepackage{amsmath,amssymb}
\usepackage{graphicx}
\usepackage{booktabs}
\usepackage{algorithm}
\usepackage{algpseudocode}
\usepackage{hyperref}
\usepackage{tikz}
\usetikzlibrary{arrows.meta,decorations.markings,shapes.geometric}
\usepackage{subcaption}
\usepackage{verbatim}
\usepackage[section]{placeins}
\usepackage{stfloats}
\usepackage[normalem]{ulem}

\makeatletter
\let\concord@dbflt\@dbflt
\def\@dbflt#1{%
  \ifstrequal{#1}{figure}
    {\@ifnextchar[{\@xdblfloat{#1}}{%
      \edef\reserved@a{\noexpand\@xdblfloat{#1}[\fps@figure]}%
      \reserved@a}}
    {\concord@dbflt{#1}}%
}
\makeatother

\def\tsc#1{\csdef{#1}{\textsc{\lowercase{#1}}\xspace}}
\tsc{WGM}
\tsc{QE}

\newcommand{\mypk}{\textsc{ConCorD}}

\newcommand{\beq}{\begin{equation}}
\newcommand{\eeq}[1]{\label{eq:#1}\end{equation}}

\newcommand{\beqa}{\begin{align}}
\newcommand{\eeqa}{\end{align}}

\begin{document}

\let\WriteBookmarks\relax
\renewcommand{\topfraction}{0.9}
\renewcommand{\dbltopfraction}{0.9}
\renewcommand{\bottomfraction}{0.8}
\renewcommand{\textfraction}{0.07}
\renewcommand{\floatpagefraction}{0.7}
\renewcommand{\dblfloatpagefraction}{0.7}
\setcounter{topnumber}{3}
\setcounter{dbltopnumber}{2}
\setcounter{totalnumber}{5}
\shorttitle{\mypk: a compiler for \textsc{Con}structing lattice QCD \textsc{Cor}relators in \textsc{D}istillation}
\shortauthors{Tanishk Shrimal, M. Padmanath}

\title[mode = title]{\mypk: A Compiler for Constructing Lattice QCD Correlators in Distillation}        

\author[1,2]{Tanishk Shrimal}[orcid=0009-0008-9568-0294]
\cormark[1]
\ead{tanishks@imsc.res.in}

\credit{Conceptualization of this study, Methodology, Software}

\affiliation[1]{organization={The Institute of Mathematical Sciences Chennai},
                addressline={IV Cross Road, CIT Campus, Taramani}, 
                city={Chennai},
                postcode={600113}, 
                state={Tamil Nadu},
                country={India}}
\affiliation[2]{organization={Homi Bhabha National Institute},
                addressline={Training School Complex, Anushaktinagar}, 
                city={Mumbai},
                postcode={400094}, 
                state={Maharashtra},
                country={India}}

\author[1,2]{M. Padmanath}[orcid=0000-0001-6877-7578]
\credit{Conceptualization of this study, Methodology, Software}

\cortext[cor1]{Corresponding author}

\begin{abstract}
We present \mypk, a Python package that computes Euclidean correlation matrices of hadron operators in the distillation framework from a quark-level description of the operator basis. The \mypk\thinspace workflow consists of two stages, \emph{compile} and \emph{run}. In the \emph{compile} stage, the user-provided operator basis is used to build a symbolic representation of the requested correlation matrix, whose Wick contractions are encoded as tensor-contraction graphs. The result is a portable plan, already reduced algebraically and optimized to a limited extent numerically, which by default depends neither on numerical data nor on the hardware. An intermediate optional stage aids in inspecting operational cost and memory requirements on a given hardware, before any numerical data are read. In the \emph{run} stage, the required numerical inputs are loaded and the requested correlation matrix is evaluated. Correlation matrices evaluated using \mypk\thinspace have been validated for basic mesons and baryons, derivative and transported meson operators, two-meson systems, diquark-antidiquark tetraquarks, baryon-meson systems, dibaryons and the triton, using external independent contraction codes and exact flavor identities. Three-point functions with flavor-changing point currents are supported and are validated with externally supplied generalized perambulators. All validation tests agree to a relative precision of about $10^{-14}$ or better, except for local mesons at rest, which agree to the precision with which the stored distillation modes are orthonormal.
\end{abstract}

\begin{keywords}
Lattice QCD \sep Distillation \sep Wick contractions \sep Correlation matrices \sep
Tensor-network contraction
\end{keywords}

\maketitle

\noindent\textbf{PROGRAM SUMMARY}\par\smallskip
\begin{small}
\noindent\emph{Program title:} \mypk{}\\
\emph{CPC Library link to program files:} TBA\\
\emph{Developer's repository link:} \url{https://github.com/Tanishk-Shrimal/ConCorD}\\
\emph{Licensing provisions:} MIT\\
\emph{Programming language:} Python (\(\ge3.11\))\\
\emph{Nature of problem:} Lattice QCD studies of excited and multi-hadron states need correlation matrices of large bases of interpolating operators. Every matrix element is a sum of Wick contractions whose number grows factorially with the number of quarks. Deriving, checking and evaluating these terms by hand is laborious and error-prone, and evaluating them one by one repeats much identical work.\\
\emph{Solution method:} The operators are written symbolically. \mypk\thinspace generates all Wick contractions with their signs, lowers each term to a tensor network of distillation kernels, perambulators and spin matrices, eliminates common subexpressions across the whole correlation matrix, chooses contraction paths, schedules the operations and plans the memory buffers. The resulting plan is executed with \texttt{PyTorch} on CPUs or GPUs for every gauge configuration and timeslice.\\
\emph{Additional comments including restrictions and unusual features:} Computation of Laplacian eigenvectors or the distillation mode space and the ordinary/generalized perambulators are beyond the scope of \mypk\thinspace and should be supplied externally. Gauge configurations, smeared if required, should also be supplied externally if displaced operators are utilized in the operator basis. All input datafiles are to be supplied in \texttt{HDF5} format, except the gauge configurations for which the \texttt{openQCD} format or in the smeared-gauge format of \texttt{chroma\_laph} and \texttt{QUDA-LapH} are supported. A single work unit runs on a single device; on a CPU node, independent work units are spread over worker processes. \mypk\thinspace dependencies include \texttt{NumPy}, \texttt{PyTorch}, \texttt{opt\_einsum}, \texttt{h5py}, \texttt{Pydantic}, \texttt{PyYAML} and \texttt{Typer}.
\end{small}

\section{Introduction}
\label{sec:introduction}

The strong coupling of Quantum ChromoDynamics (QCD) at low energies, where hadrons emerge as the relevant effective degrees of freedom, precludes the use of perturbative approaches. Consequently, hadron physics investigations in the low energy regime rely on nonperturbative approaches such as those using the lattice QCD framework. Generation of lattice QCD ensembles through Monte-Carlo simulations and estimation of fermion matrix determinant as well as inversions of the fermion matrix consume substantial fraction of the computer time in lattice QCD calculation and have been extensively discussed and investigated since the inception of lattice QCD \cite{Duane:1987de,Babich:2010qb,Clark:2009wm,openQCD}. Physical observables in hadron-physics investigations are extracted from Euclidean correlation functions. In the early years of lattice QCD, these calculations primarily targeted ground-state hadron properties and were comparatively inexpensive. The discovery and study of numerous heavy and excited hadrons over the past several decades, however, necessitated increasingly sophisticated spectroscopy calculations, notably those employing Euclidean correlation matrices and variational analyses to determine the finite-volume spectrum and its properties. As the size and complexity of these correlation matrices increase, the associated operator construction, Wick contractions, and numerical evaluation become increasingly demanding, resulting in substantial rise in computational costs.

Modern day hadron spectroscopy proceeds through evaluation of two point Euclidean correlation matrices 
\beq 
C_{ij}(t-t_0)=\left\langle\mathcal O_i(t)\,\mathcal O_j^{\dagger}(t_0)\right\rangle
\eeq{2ptCfn}
for a basis of interpolators/operators $\{O_i\}$ that carry the quantum numbers of the desired hadronic state, where $t$ and $t_0$ refer to the sink and the source timeslices. In this context, the new challenges include a systematic construction of interpolating operators that enter the evaluation of correlation matrices, handling the allowed Wick contractions for every combination of interpolating operators, keeping track of the fermionic signs with permutations, flavor, spin, color and spatial indices faithfully, etc. The amount of bookkeeping therefore grows as one moves from meson operators to baryons, multi-hadron operators and nuclei. Additional conversions and reductions need to be made to express the correlation matrices in terms of the objects defining operator structures as well as the fermion matrix inverses. Note that different matrix elements carry similarities and share quark-propagation data, operator-dependent building blocks, and often portions of the resulting contraction expressions. This observation suggests opportunities to identify repeated (sub)objects, input/output operations and computations, reducing the computational redundancy. While a systematic road-map has already been laid out for such excited hadron spectroscopic investigations in the literature~(cf.~\cite{Dudek:2009qf,Edwards:2011jj,Dudek:2014qha,Padmanath:2014arv,Padmanath:2018zqw,Briceno:2017qmb,Green:2021qol,BaryonScatteringBaSc:2023zvt,Green:2026jgv}), performing them manually for every new channel being studied requires substantial implementation effort and can introduce bookkeeping errors.

There already exist tools to handle the operator construction and correlation matrix evaluation. Tools such as Multi-Hadron Interpolators (\texttt{MHI})~\cite{PhysRevD.109.094516} and \texttt{OpTion}~\cite{Yan:2025op} provide a systematic construction of multi-hadron lattice operators projected to finite-volume irreps, total lab frame momentum, spin, and internal quantum numbers. The QCT package \cite{Djukanovic:2016spv} performs the Wick contractions of a given set of hadronic currents/operators expressed in terms of (anti)quark fields and writes the results as \texttt{QDP++}~\cite{Edwards:2004sx} expressions for a \texttt{C++} program, without further optimization for computational efficiency. The \texttt{Redstar} framework~\cite{10.1145/3592979.3593409} expresses each element in correlation matrices as a sum of graphs (Wick contractions) and uses graph-contraction algorithms to reduce redundant evaluations and to organize optimized correlator evaluations in a distillation-based calculational setup \cite{Peardon:2009gh} using tensor contractions for execution on CPUs and GPUs. Building on \texttt{Redstar}, Ref.~{\cite{Selvitopi:2025sched}} orders the contractions of many graphs so as to reduce the peak memory on a GPU and the data traffic between the host and the device. Another recent algorithmic work explored determinant representations of Wick contractions and methods for identifying smaller sets of determinant terms that are sufficient for evaluating nuclear correlation functions~\cite{PhysRevD.110.114505}.

The emphasis of the \mypk\thinspace package is on having a complete automated transformation from a symbolic quark-level operator description to a numerical calculation within the distillation prescription~\cite{Peardon:2009gh} as a sequence of distinct computational stages. Once the user provides the symbolic definition of an operator basis in terms of quark and gauge fields, the operator representations, Wick contractions, automated translation to backend-independent physics objects, distillation lowering, execution planning, and numerical evaluation are organized as distinct layers of the software. This layered organization reduces the manual intervention needed in correlator-specific algebra and implementation, while maintaining a clear separation between the physical definition of an operator and the details of its numerical evaluation. It also makes the intermediate algebraic transformations explicit and amenable to inspection and validation, while allowing numerical execution strategies to evolve and adapt independent of the underlying physics description of the interpolating operators.

The paper is organized as follows. Section~\ref{sec:background} reviews the distillation framework and defines the relevant objects referred to later on in the paper. Section~\ref{sec:overview} provides an overview of the \mypk\thinspace workflow and its scope. Section~\ref{sec:symbolic} describes the symbolic representation of interpolating operators. In Sec.~\ref{sec:wick}, we discuss building the tensor network representation of all involved Wick contractions in the correlation matrix to be evaluated. Section~\ref{sec:distillation_lowering} elaborates how the Wick contraction networks are transformed into the distillation language. The compilation process involved in building a portable optimized plan, including its efficacy with the provided hardware, for the actual numerical evaluation is summarized in Sec.~\ref{sec:execution}. Sections~\ref{sec:validation} and~\ref{sec:performance} present the numerical validation and the performance evaluation. Section~\ref{sec:software} describes the software availability and its usage. In Sec.~\ref{sec:limitations}, we list a number of limitations and future developments associated with the package. We summarize in Sec.~\ref{sec:summary}.

\section{Distillation framework}
\label{sec:background}
\mypk\thinspace facilitates the computation of Euclidean correlation matrices in the distillation framework \cite{Peardon:2009gh}, serving itself as a generalized automated platform in computation of general multi-hadron correlation matrices in finite-volume, with the assumption that the relevant lattice numerical data and the operator basis details are provided by the user. We begin this section by briefly discussing numerical objects relevant for the distillation program for completeness. In the conventional non-distillation setup, the measurements involve computation of correlation matrices given in Eq.~\ref{eq:2ptCfn} for a given operator basis and varying values of $t$ and $t_0$. These involve appropriate tensor index contractions between spatial-color-spin indices of the hadron interpolators with those in the quark line objects referred to as the quark propagators.

The computational cost of such tensor index contractions spanning through the space-color-spin index values ($\mathcal{O}(10^6)$ for a $40^3$ spatial lattice) can become prohibitively expensive when the system being studied is large or complex such as multi-hadron operators, tetraquarks, pentaquarks, nuclei, etc. In the distillation program~\cite{Peardon:2009gh}, the combined space-color index is replaced by a single index, which runs over an appropriately chosen subset of orthonormal basis $\{v^{(i)}(t)\}$. Such a replacement can be facilitated through quark smearing program defined in terms of $\{v^{(i)}(t)\}$ as
\begin{align}
\tilde q(t)=\Box(t)q(t), &\qquad
\tilde{\bar q}(t)=\bar q(t)\Box(t), \nonumber \\
\mbox{~~where ~~~} \Box(t)&=V(t)V^\dagger(t) = \sum_{i=1}^{N_v}|v^{(i)}(t)\rangle\langle v^{(i)}(t)|,  \nonumber \\
\mbox{~and ~~~} V(t)&=\left(v^{(1)}(t),v^{(2)}(t),\ldots,v^{(N_v)}(t)\right). \label{eq:distill}
\end{align}
The $\Box(t)$ operator acts as a low pass filter or a projection operator onto the space spanned by the $N_v$ vectors. If the orthonormal basis is chosen to be the eigenmodes of the discretized gauge-covariant Laplacian, this procedure amounts to building a low mode projector of the Laplacian operator with a Heaviside step function, which is often referred to as  Laplacian-Heaviside (\texttt{LapH}) smearing~\cite{Morningstar:2011ka}. \mypk\thinspace evaluates correlations with this exact low mode projector. Weighted profiles of the modes and stochastic estimators are possible extensions (Sec.~\ref{sec:limitations}). Inserting such projection operators for the quark fields in the conventional correlation functions, followed by regrouping, leads to the new objects
\begin{align}
\tau_f(t,t_0) &= V^\dagger(t) M_f^{-1}(t,t_0) V(t_0) \label{eq:peram}\\
\Phi_{\mathcal{O}}(t)&= V^\dagger(t)\,\mathcal{O}\,V(t). \label{eq:kernel} 
\end{align}
referred to as the perambulators and the elementals (kernels), which carry the information of the quark propagator $M_f^{-1}(t,t_0)$ and of the hadronic operator $\mathcal O$, respectively. The objects $\tau_f(t,t_0)$ and $\Phi_{\mathcal{O}}(t)$ carry indices running over the $N_v$ modes, typically a few tens to a few hundreds, facilitating contractions of tensor indices with reduced computational cost. This procedure was not only found to reduce the computational cost requirement involved in tensor contractions (at the level of construction of correlation matrices), but also found to provide access to all-to-all quark propagators, to source momentum projection in hadron operator, multi-hadron operators, etc. The expensive quark-propagation information can be prepared independently of a particular spectroscopy operator basis, while changes to momenta, spin structures, or spatial constructions modify the operator-dependent factors. Within the distillation framework, the correlation matrix elements can be written in a compact form as
\begin{equation}
C_{ij}(t,t_0)= \sum_{\alpha} s_{\alpha}\,c_{\alpha}\,\operatorname{Contr}_{\alpha}\left[\{\Phi\},\{\tau\}\right],
\label{eq:distillation_contraction_schematic}
\end{equation}
where $\alpha$ runs over the Wick-contractions ``$\operatorname{Contr}_{\alpha}\left[...\right]$'' and $s_{\alpha}=\pm1$ and $c_{\alpha}$ are the associated fermionic sign and algebraic coefficient.

\section{Overview and scope of \mypk \label{sec:overview}}
In \mypk, the user-provided interpolators determine the Wick contractions and their representation in the distillation space, as illustrated schematically in Eq.~\ref{eq:distillation_contraction_schematic}, while the subsequent numerical procedures, including tensor contractions, redundancy optimizations to avoid repetitive evaluations, memory (de)allocation, and data reuse, are handled by the \mypk\thinspace backend. Every resource of the compiled plan keeps its physics description (flavor, quark fields, momenta, spin structure), and their corresponding symbolic representations can be retrieved at any time, to facilitate user-end as well as backend verifications and troubleshooting, while the numerical tensor objects and operations remain blind to the physics that is aimed at. The correlation matrix evaluation is treated within the \mypk\thinspace package as a single structured computational problem,  instead of compiling every matrix element as an independent numerical program. This helps in identifying similar algebraic expressions as well as redundant mathematical work across the full calculation and thus providing efficient choices for tensor contraction ordering and data reuse, enabling an optimized workflow. 

\begin{figure}[pos=!hb]
\centering
\resizebox{0.90\columnwidth}{!}{%
\begin{tikzpicture}[
    font=\small,
    stage/.style={draw, rounded corners=2pt, align=center, text width=#1,
                  inner sep=3.5pt, minimum height=0.85cm, fill=white},
    stage/.default=6.3cm,
    io/.style={draw, align=center, inner sep=3.5pt, minimum height=0.8cm, fill=black!7},
    arr/.style={-{Latex[length=2mm]}, thick},
    frame/.style={draw=black!50, dashed, rounded corners=5pt},
    side/.style={rotate=90, font=\footnotesize\itshape, text=black!75},
  ]
  \node[io, text width=6.3cm] (in) at (0,0)
    {Interpolating operators (Python) and job specification (\texttt{YAML})};

  \node[stage] (c1) at (0,-1.45)
    {Expansion of operator combinations;\\ exact Wick contraction (pairings, signs)};
  \node[stage] (c2) at (0,-2.65)
    {Distillation lowering: spatial kernels,\\ propagation resources, spin tensors};
  \node[stage] (c3) at (0,-3.85)
    {Global tensor network: shared resources,\\ canonical components, term aggregation};
  \node[stage] (c4) at (0,-5.05)
    {Contraction order, shared intermediates,\\ schedule and buffer plan};
  \draw[frame] (-3.5,-0.80) rectangle (3.5,-5.65);
  \node[side] at (-3.8,-3.225) {compile: \texttt{concord plan}};

  \node[io, text width=2.8cm] (plan) at (-1.85,-6.40)
    {Portable plan\\ (\texttt{.concord})};
  \node[io, text width=2.8cm] (data) at (1.95,-6.40)
    {Lattice data:\\ $V(t)$, $\tau_f(t,t_0)$,\\ configuration $U$};

  \node[io, dashed, text width=2.7cm, inner sep=3pt, font=\footnotesize] (insp) at (0.15,-7.65)
    {optional:\\ \texttt{concord inspect}\\ cost, hardware check};

  \node[stage=4.6cm] (r1) at (-1.05,-9.15)
    {Plan validation; device choice\\ and time batching;\\ worker processes};
  \node[stage=4.6cm] (r2) at (-1.05,-10.55)
    {Data binding; kernels and\\ propagators materialized};
  \node[stage=4.6cm] (r3) at (-1.05,-11.75)
    {Execution of the plan for\\ each configuration and time};
  \draw[frame] (-3.5,-8.30) rectangle (3.5,-12.45);
  \node[side] at (-3.8,-10.375) {run: \texttt{concord run}};

  \node[io, text width=6.3cm] (out) at (0,-13.20)
    {Correlation functions (\texttt{HDF5});\\ optional run record and report};

  \foreach \n/\k in {c1/1,c2/2,c3/3,c4/4,r1/5,r2/6,r3/7}
    \node[circle, draw, fill=white, inner sep=0.8pt, font=\tiny\bfseries] at (\n.west) {\k};

  \draw[arr] (in) -- (c1);
  \draw[arr] (c1) -- (c2);
  \draw[arr] (c2) -- (c3);
  \draw[arr] (c3) -- (c4);
  \draw[arr] (c4.south -| plan.north) -- (plan.north);
  \draw[arr] (plan.south) -- (plan.south |- r1.north);
  \draw[arr, dashed] (plan.south |- insp.west) -- (insp.west);
  \draw[arr] (data.south) -- (data.south |- r2.east) -- (r2.east);
  \draw[arr] (r1) -- (r2);
  \draw[arr] (r2) -- (r3);
  \draw[arr] (r3.south) -- (r3.south |- out.north);
\end{tikzpicture}%
}
\caption{\mypk\thinspace workflow. The \emph{compile} stage depends only on the symbolic operators and the compile settings, including the declared eigenvector format, and produces a portable, hardware-independent plan, whose cost can optionally be inspected and checked against a machine before any data are read. Lattice data (Laplacian eigenvectors $V(t)$, perambulators $\tau_f(t,t_0)$, and the gauge configuration $U$) and the available hardware details enter only at \emph{run} stage. The circled numbers on the left of each block label the layers of the software.}
\label{fig:workflow}
\end{figure}
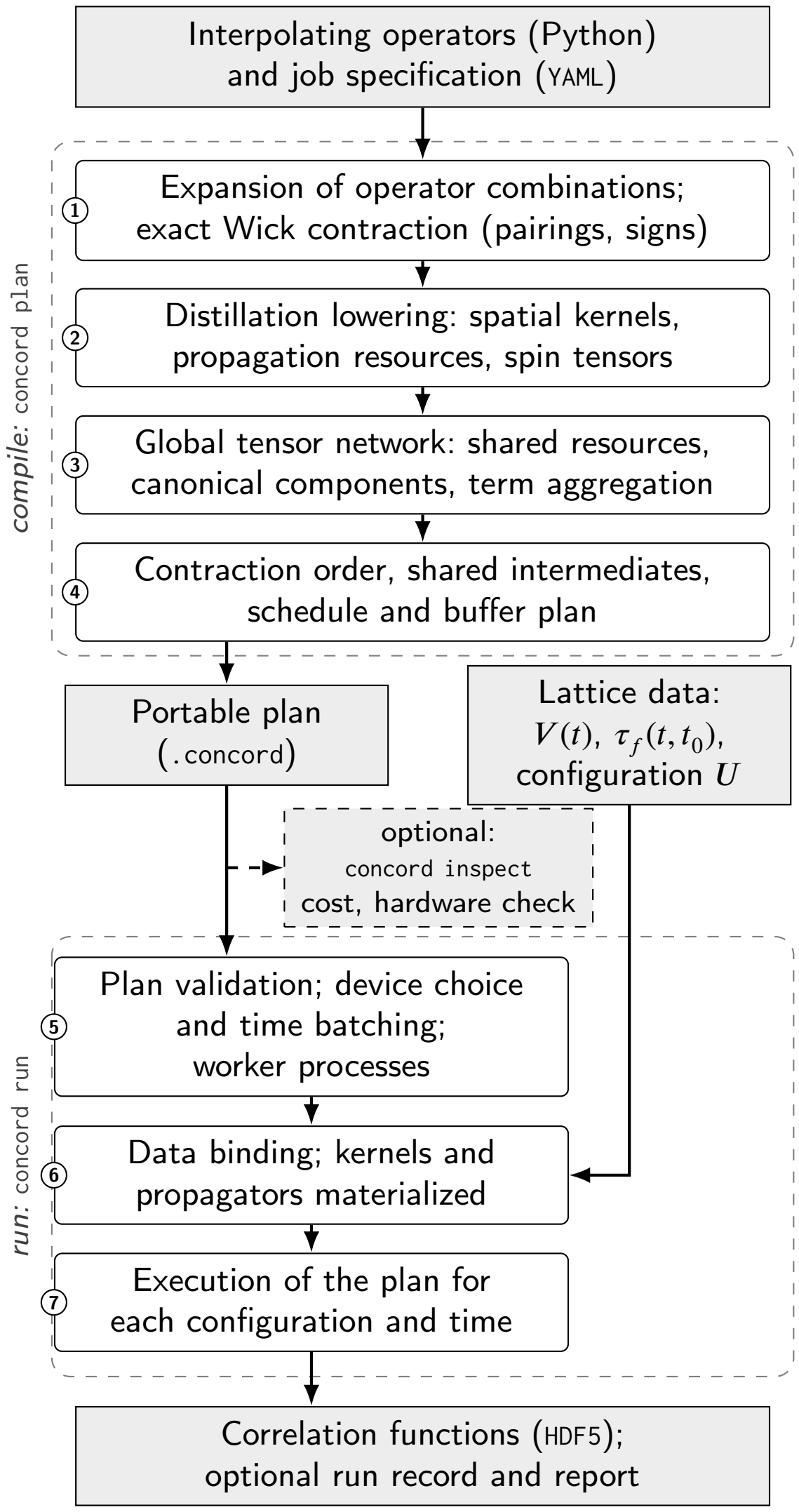

We present the \mypk\thinspace workflow in Fig.~\ref{fig:workflow}. The user provides the interpolator basis in symbolic form (Sec.~\ref{sec:symbolic}) together with a job specification that fixes the lattice dimensions, the number of distillation modes, the spin basis, the data locations, and the source and sink times. The calculation then proceeds in two stages, \emph{compile} and \emph{run}. The \emph{compile} stage uses only the symbolic operators and the compile settings, including the declared eigenvector format. Linear combinations of operators are expanded into basis correlators, and all quark-line pairings (Wick contractions), their fermionic signs, and their coefficients are generated for every element of the correlation matrix (Sec.~\ref{sec:wick}). The resulting expressions are lowered to the distillation framework, where each Wick term becomes a tensor network of operator-dependent spatial kernels, propagation (perambulator) resources, and spin tensors (Sec.~\ref{sec:distillation_lowering}). All tensor networks of the requested correlation matrix are then treated as one global contraction problem: identical resources and identical contraction networks are shared across Wick terms and matrix elements, exactly equal terms are combined, a contraction order is chosen, shared intermediates are identified, and the operations are scheduled together with a plan for reusing memory buffers (Sec.~\ref{sec:execution}). The result is saved as a portable plan, a \texttt{.concord} file, that depends neither on the lattice data, apart from the number of modes and the declared eigenvector format, nor on the hardware. Different plans can be built based on path-search strategies, the efficacy of which can be assessed using the \texttt{concord inspect} feature, which returns a report on the cost of the plans (operations and memory) before any data are read. This allows the user to compare different plans before the actual numerical execution and to decide on a plan that fits within the available hardware. 

The second stage is the \emph{run} stage, where \mypk\thinspace first checks the saved plan against the current operator definitions and the compile settings of the job specification. Following this, \mypk\thinspace reads the Laplacian eigenvectors $\{v^{(i)}(t)\}$, the perambulators $\tau_f(t,t_0)$, and the gauge configuration $\{U_{\mu}(\bar x, t)\}$; builds the required kernels and propagation tensors; and executes the plan for every gauge configuration and source time. By default, \mypk\thinspace chooses at run time how many timeslices are evaluated together and, on a CPU node, the number of worker processes and their threads (\texttt{run.time\_batch\_size} and \texttt{run.workers} set to \texttt{auto}); the user can also fix them. With \texttt{run.device: auto}, \mypk\thinspace also chooses between the CPU and a GPU. The run-time choices are not stored in the plan, so the same plan can be run with different choices on different machines. 

All optimizations, at the algebraic, numerical, and execution stages, are handled by the \mypk\thinspace backend while preserving the underlying physics information. This shields non-expert users from the complexities of optimization, while still giving expert users full freedom to inspect the practical implementation and numerical performance. \mypk\thinspace does not perform gauge smearing, and thus any required smearing of the gauge configurations must be done externally beforehand. The correlation functions are written to an \texttt{HDF5} file; a record of the run and a human-readable report can be produced on request.\label{sec:numerical_execution}

Now we present the scope of the \mypk\thinspace package. The present implementation can serve in optimized distillation-based correlation matrix evaluations and works on both CPU as well as CUDA GPU resources. The package assumes externally supplied operator basis, lattice QCD configuration, associated Laplacian eigenvectors, and the quark perambulators. Operator constructions such as those provided by the packages \texttt{MHI} and \texttt{OpTion}~\cite{PhysRevD.109.094516,Yan:2025op}, generation of the lattice QCD configurations \cite{openQCD, MILC} and evaluation of Laplacian eigenvectors, and perambulators \cite{quada_laph} are outside \mypk's scope. The package can inspect the available resources such as CPUs, GPUs, memory, and can make educated decisions on efficient execution, however it does not support multi-GPU and MPI based multi-CPU evaluations of a single work unit. Independent work units (configurations and timeslices) are already spread over the cores of one node by worker processes that share one loaded plan, and can be distributed over devices or nodes as separate jobs that reuse one compiled plan, whereas distributing a single work unit is future work. These are limited execution boundaries rather than restrictions on the symbolic definition of the correlator.

At the physics level, the correlators that have been validated against independent codes and exact identities are discussed in Sec.~\ref{sec:validation}. Beyond these explicitly verified cases, the package has additional capabilities that include, but are not limited to, larger multi-hadron correlation matrices,  derivative and displaced operators with other spin structures or other displaced baryon legs, and correlation matrices involving disconnected quark loops. Three-point correlation functions with a flavor-changing current, built from generalized perambulators~\cite{Egerer:2018xgu}, have been verified against an independent evaluation on synthetic data. Flavor-diagonal local currents, which also produce a quark loop at the current, are not yet supported. Supporting them is a natural extension, once estimates of the quark loops with the corresponding local currents are provided externally to \mypk.

\section{Symbolic Representation of Interpolating Operators \label{sec:symbolic}}

The \mypk\thinspace package takes in the interpolator basis as a list of ordered algebraic expressions built out of objects representing (anti)quark fields, gauge fields, spin and color structures. At this stage of interpolator definition, the representation is independent of distillation tensors (perambulators and operator kernels/elementals), data layouts, contraction paths, and execution backends. 

Quark and antiquark fields of flavor f $\in\{u,d,s,c,b,t\}$, or a generic light flavor $l$, are represented symbolically as 
\begin{equation}
    q_{\rm f}(x)_{a\alpha}\equiv\mbox{\texttt{f("x", "a", "alpha")}} 
\end{equation} 
and 
\begin{equation}
\bar q_f(x)_{a\alpha}\equiv \mbox{\texttt{fbar("x", "a", "alpha")}},
\end{equation} 
where ($x$, \texttt{x}), ($a$, \texttt{a}), and ($\alpha$, \texttt{alpha}) denote position, color, and Dirac spin indices, respectively. The explicit flavor information and the quark/antiquark identification in the fermion object names aid in identifying flavor compatibility and in keeping track of fermionic permutation signatures in Wick contractions at the \emph{compile} stage. The names chosen for summed spin, color and position indices carry no meaning beyond the connectivity they define (Sec.~\ref{sec:wick}). The package assumes the following convention for the index labels. When a label is first introduced, its index type (spin, color or space) is fixed; attempting to reuse the same label for a different type will be rejected. Reusing the label within the same type defines the intended tensor connectivity of the operator.

Until the backend stage of numerical evaluation, no numerical conventions are assumed for the Dirac gamma structures. At the front end, the user may provide usual gamma matrices to represent the operator spin structures, examples of which will be given later in this section. Numerically, \texttt{gammaX}, \texttt{gammaY}, \texttt{gammaZ} and \texttt{gammaT} are the DeGrand--Rossi matrices of \texttt{Chroma}/\texttt{QDP++}, or their Dirac--Pauli transforms when the user chooses that basis using the \texttt{compile.spin\_basis} option. On the color index, the package supports the Kronecker delta $\delta_{ab}$ for the meson-like objects and the antisymmetric Levi-Civita tensor $\epsilon_{abc}$ for the baryon-like operators. Spin and color indices follow the usual Einstein repeated index contraction convention within a symbolic expression. A single spin index occurrence remains open and can be used to represent the uncontracted Dirac spin index, such as in baryon operators. A single color index is allowed by the representation. The function \texttt{validate(op)} reports it as a warning, because the resulting operator is gauge dependent. The same function also reports a spin or color index appearing more than twice as an error. The compile stage does not check color-index multiplicities; this check is left to the user through \texttt{validate(op)}. 

Position indices identify the spatial locality associated with fermionic fields within an expression and are therefore not subject to the Einstein summation convention. Two or more fermionic fields carrying the same position label belong to the same locality group, whereas distinct labels can be used to represent multi-local or more general composite operators. This convention helps in assigning a definite three-momentum to each locality group as 
\begin{equation}
 x_r \longmapsto \mathbf p_r,
 \qquad
 \mathbf P = \sum_{r=1}^{n_l} \mathbf p_r ,
\label{eq:operator_momentum_groups}
\end{equation}
where $n_l$, $\mathbf p_r$ and $\mathbf P$ are the number of locality groups within, the three-momentum of locality group labelled $r$ and the total three-momentum carried by the full symbolic expression. In this sense, the position labels represent the momentum orientations within the symbolic expression. Spatial transformations such as pure translations, gauge-covariant parallel transport, covariant derivatives, including the directional information on the fermionic fields are also supported by dedicated objects: \texttt{shift}, \texttt{transport}, and \texttt{derivative}. 

Large symbolic expressions that are representative of hadronic operators can thus be constructed as products of elementary fermion field objects, with the factor ordering retained explicitly to ensure a faithful evaluation of the permutation signs during Wick contractions. Momentum information and linear combinations of such expressions can then be assigned to build composite interpolating operators, with further freedom to provide additional metadata. Operators that differ only in the names of their summed indices are recognized as identical (Sec.~\ref{sec:wick}). The sink and source operator bases may be provided as separate ordered objects by the user. When an explicit source basis is not supplied, the source operator basis is constructed as the Hermitian conjugate of the sink operator basis. 

Now we provide various examples of input operator structures from which more complicated operators can be further built. A generic local meson-like operator is given by a quark bilinear $\mathcal O_M(x)=\bar q_{f_1}(x)_{a\alpha}\Gamma_{\alpha\beta} q_{f_2}(x)_{a\beta}$. For the isovector pseudoscalar pion operator, this is given by $\mathcal O_{\pi}(x) = \bar d(x)\,\gamma_5\,u(x)$, where the spin-color indices are implicit. The input expression for this operator projected to zero momentum at the sink is given by 
\begin{verbatim}
pionsnk = Operator( dbar("x", "a", "alpha")
                *gamma5("alpha", "beta")
                *u("x", "a", "beta"),
                 momentum=(0, 0, 0),)
\end{verbatim}
The corresponding source operator can be built from the operator definition at the sink as \texttt{pionsrc = pionsnk.dagger()}.

The conventionally used local baryon-like operator structure is given by diquark-passenger quark form given by $\mathcal O_B(x)_\delta=\epsilon_{abc}[q_1(x)_{a\alpha}\Gamma_{\alpha\beta}q_2(x)_{b\beta}]q_3(x)_{c\delta}$, where $\Gamma$ carries the spin structure of the diquark, whereas the $\delta$ is the Dirac spin index of the passenger quark $q_3$ as well as the baryon. The simplest such quark trilinear operator for a proton is built by the assignments $q_1\rightarrow u$, $q_3\rightarrow u$, $q_2\rightarrow d$ and $\Gamma\rightarrow C\gamma_5$. The symbolic representation of such a proton operator is given by 
\begin{verbatim}
proton = Operator( epsilon("a", "b", "c")
                  *cgamma5("alpha", "beta")
                  *u("x", "a", "alpha")
                  *d("x", "b", "beta")
                  *u("x", "c", "delta"),
                   momentum=(0, 0, 0),)
\end{verbatim}
One could also build other proton operator with a nontrivial gamma structures on the passenger quark in a similar fashion. \mypk\thinspace also comes with a convenience constructor for such diquark-passenger quark form of baryon-like operators, using which a proton operator can be built as 
\begin{verbatim}
proton = build_baryon_operator( 
         ["u", "d", "u"], "Cg5",
         momentum=(0, 0, 0),)
\end{verbatim}
More complex local hadron operators can further be constructed from these elementary objects and constructions. A local scalar diquark-antidiquark structure representing a compact tetraquark operator can be represented as\\
\begin{verbatim}
dqadq_local = Operator(  epsilon("a", "b", "e") 
                       * u("x", "a", "alpha") 
                       * cgamma5("alpha", "beta") 
                       * d("x", "b", "beta")
                       * epsilon("c", "d", "e")
                       * ubar("x", "c", "gamma") 
                       * cgamma5("gamma", "delta") 
                       * dbar("x", "d", "delta"),
                       momentum=(0, 0, 0),)
\end{verbatim}
Another example of a simple tetraquark operator is that of a local two-meson structure, which can be represented as 
\begin{verbatim}
two_meson_local = Operator(  dbar("x", "a", "alpha")
                           * gamma5("alpha", "beta")
                           * u("x", "a", "beta")
                           * ubar("x", "b", "gamma")
                           * gamma5("gamma", "delta")
                           * d("x", "b", "delta"),
                           momentum=(0, 0, 0),)
\end{verbatim}
Note that in the above construction there are two color singlet quark bilinears forming two independent mesons both featuring in the same locality group leading to a single momentum assignment for the entire structure. Multiplying two separately defined \texttt{Operator} objects that use the same position label to define a new operator is not supported.

Spatially extended operators are built by passing the relevant elementary field as the argument of one of the operations \texttt{shift} $S_{\pm i}$, \texttt{transport} $T_{\pm i}$ and \texttt{derivative} $\nabla_{\pm i}$, which act on that field only. For a spatial direction $i$, they are defined as
\begin{align}
S_{\pm i}\,\psi(x) &= \psi(x\pm\hat\imath), \nonumber\\
T_{+i}\,\psi(x) &= U_i(x)\,\psi(x+\hat\imath), \nonumber\\
T_{-i}\,\psi(x) &= U_i^\dagger(x-\hat\imath)\,\psi(x-\hat\imath), \label{eq:transport_conventions}\\
\nabla_{\pm i}\,\psi &= \pm\tfrac12\left(T_{+i}-T_{-i}\right)\psi, \nonumber
\end{align}
with periodic spatial boundaries and no lattice-spacing factor. A transport path is listed from the field position towards the displaced endpoint, so that \texttt{transport(q, "+x", "+y")} realizes 
\begin{equation}
T_{+x}T_{+y}\,q(x)=U_x(x)\,U_y(x+\hat x)\,q(x+\hat x+\hat y),
\end{equation} 
and nested operations are applied in their symbolic order, e.g.\ \texttt{derivative(transport(q, "+y"), "+x")} is $\nabla_{+x}T_{+y}\,q$. When a source operator is obtained with \texttt{dagger()}, the operation stays on the conjugated field: \texttt{transport(u(...), "+x")} becomes \texttt{transport(ubar(...), "+x")}, i.e.
\begin{equation}
    [T_{+x}u(x)]^\dagger\gamma_4=\bar u(x+\hat x)\,U_x^\dagger(x).
\end{equation}

Below are three examples of simple meson operators (projected to zero momentum) with nontrivial spatial structures. The first one, $\mathcal{O}_{M_1}=\bar d(x)\gamma_5\nabla_{+x}u(x)$, contains the symmetric derivative in the $x$ direction, and the corresponding symbolic expression is
\begin{verbatim}
mesondx = Operator(  dbar("x", "a", "al")
                   * gamma5("al", "be")
                   * derivative(u("x", "a", "be"),"+x",),
                   momentum=(0, 0, 0),)
\end{verbatim}
\tikzset{
  site/.style={circle, draw, fill=white, inner sep=1.4pt},
  qsite/.style={circle, draw, fill=black, inner sep=1.4pt},
  link/.style={very thick, -{Latex[length=1.8mm]}},
  lbl/.style={font=\scriptsize},
}
\begin{center}
\begin{tikzpicture}
  \draw[black!35] (-1.6,0) -- (1.6,0);
  \node[qsite] (m) at (-1.2,0) {}; \node[site] (o) at (0,0) {}; \node[qsite] (p) at (1.2,0) {};
  \draw[link] (o) -- node[lbl, above] {$U_x(x)$} (p);
  \draw[link] (o) -- node[lbl, above] {$U_x^\dagger(x-\hat x)$} (m);
  \node[lbl, below=2pt] at (o) {$\bar d(x)$};
  \node[lbl, below=2pt] at (p) {$+\tfrac12\,u$};
  \node[lbl, below=2pt] at (m) {$-\tfrac12\,u$};
\end{tikzpicture}
\end{center}
The second operator contains a forward gauge-covariant transport of the $u$-quark field along the $x$ direction, $\mathcal{O}_{M_2}=\bar d(x)\gamma_5T_{+x}u(x)=\bar d(x)\gamma_5U_x(x)\,u(x+\hat x)$, and the associated symbolic expression is
\begin{verbatim}
mesontx = Operator(  dbar("x", "a", "al")
                   * gamma5("al", "be")
                   * transport(u("x", "a", "be"),"+x",),
                   momentum=(0, 0, 0),)
\end{verbatim}
\tikzset{
  site/.style={circle, draw, fill=white, inner sep=1.4pt},
  qsite/.style={circle, draw, fill=black, inner sep=1.4pt},
  link/.style={very thick, -{Latex[length=1.8mm]}},
  lbl/.style={font=\scriptsize},
}
\begin{center}
\begin{tikzpicture}
  \node[site] (o) at (0,0) {}; \node[qsite] (r) at (1.3,0) {};
  \draw[link] (o) -- node[lbl, below] {$U_x(x)$} (r);
  \node[lbl, below=2pt] at (o) {$\bar d(x)$};
  \node[lbl, right=2pt] at (r) {$u(x{+}\hat x)$};
\end{tikzpicture}
\end{center}
The third operator generalizes the second to an ordered two-link path, first along $x$ and then along $y$, $\mathcal{O}_{M_3}=\bar d(x)\gamma_5T_{+x}T_{+y}u(x)=\bar d(x)\gamma_5U_x(x)\,U_y(x+\hat x)\,u(x+\hat x+\hat y)$, and the associated symbolic expression is
\begin{verbatim}
mesontxy = Operator(  dbar("x", "a", "al")
                    * gamma5("al", "be")
                    * transport(u("x", "a", "be"),"+x","+y",),
                    momentum=(0, 0, 0),)
\end{verbatim}
\tikzset{
  site/.style={circle, draw, fill=white, inner sep=1.4pt},
  qsite/.style={circle, draw, fill=black, inner sep=1.4pt},
  link/.style={very thick, -{Latex[length=1.8mm]}},
  lbl/.style={font=\scriptsize},
}
\begin{center}
\begin{tikzpicture}
  \node[site] (o) at (0,0) {}; \node[site] (r) at (1.3,0) {}; \node[qsite] (ru) at (1.3,1.0) {};
  \draw[link] (o) -- node[lbl, below] {$U_x(x)$} (r);
  \draw[link] (r) -- node[lbl, right] {$U_y(x{+}\hat x)$} (ru);
  \node[lbl, below=2pt] at (o) {$\bar d(x)$};
  \node[lbl, left=2pt] at (ru) {$u(x{+}\hat x{+}\hat y)$};
\end{tikzpicture}
\end{center}
The directions passed to \texttt{transport} list the path walked from the field position, so the gauge links appear in the listed order, $U_x(x)\,U_y(x+\hat x)$; in the operator product $T_{+x}T_{+y}$ the last listed hop acts on the field first. Explicit gauge links and other color structures can also be introduced as factors in the symbolic expression, although standalone \texttt{GaugeLink} tensors are not currently executable by the numerical backend.

\FloatBarrier

Composite operators such as bilocal or multilocal operators could be built as products of simpler structures. Let $H_1(\mathbf p_1)$ and $H_2(\mathbf p_2)$ be independently defined operators with separate momentum assignments; their product is given by $\mathcal O_{12} = H_1(\mathbf p_1)\,H_2(\mathbf p_2)$, where $\mathbf P=\mathbf p_1+\mathbf p_2$ is the total three-momentum of the operator $\mathcal O_{12}$. In the symbolic form, individual expressions for mesons can be concatenated to form the composite operator expression either explicitly or implicitly. In the latter strategy, one defines individual mesons as  
\begin{verbatim}
meson1 = Operator(  dbar("x", "a", "alpha")
                  * gamma5("alpha", "beta")
                  * u("x", "a", "beta"),
                  momentum=(1, 0, 0),)
\end{verbatim}
and 
\begin{verbatim}
meson2 = Operator(  ubar("y", "b", "gamma")
                  * gamma5("gamma", "delta")
                  * d("y", "b", "delta"),
                  momentum=(-1, 0, 0),)
\end{verbatim}
Following these definitions, the bilocal two-meson operator, where the individual mesons are projected to definite momentum can be represented as 
\begin{verbatim}
two_meson_bilocal = meson1 * meson2
\end{verbatim}
Alternatively, one can define the same explicitly as 
\begin{verbatim}
two_meson_bilocal = Operator(  dbar("x", "a", "alpha")
                             * gamma5("alpha", "beta")
                             * u("x", "a", "beta")
                             * ubar("y", "b", "gamma")
                             * gamma5("gamma", "delta")
                             * d("y", "b", "delta"),
                             momenta={"x": (1, 0, 0),
                                    "y": (-1, 0, 0),},)
\end{verbatim}
where the individual mesons fall into separate locality groups represented by $x$ and $y$, which have momentum assignments $\mathbf p_x=(1, 0, 0)$ and $\mathbf p_y=(-1, 0, 0)$, respectively. Similarly, two-baryon operators can be built as products of single-baryon operators with a spin coupling. The $I_3=0$ component of the isovector scalar dibaryon operator is given by
\begin{verbatim}
C = spin_matrix("Cg5", "s1", "s2")
nn0 = (neutron("x", open_spin="s1")
         * proton("y", open_spin="s2") * C
     + proton("x", open_spin="s1")
         * neutron("y", open_spin="s2") * C) / sqrt(2)
\end{verbatim}
Here \texttt{proton} and \texttt{neutron} are short helper functions written by the user, not part of the package, that return the local nucleon operator above at the given position, with the open spin index named by \texttt{open\_spin} and with color and spin labels of their own. Three-nucleon operators can be constructed in a similar fashion as for two nucleons. For example, the triton nucleus operator can be written as 
\begin{verbatim}
triton =   proton("x", open_spin="sN") 
         * neutron("y", open_spin="s1") 
         * neutron("z", open_spin="s2") 
         * spin_matrix("Cg5", "s1", "s2")
\end{verbatim}

Another important feature supported by the package is the construction of complete composite operators, which are typically linear combinations of various expressions $\mathcal O =\sum_i c_i\,\mathcal O_i$. Addition, subtraction, scalar multiplication, and products of simpler expressions are also accepted as user-defined operator expressions, facilitating such composite operator expression representations. Hermitian conjugation acts on the full symbolic construction, combined with a reversed sign for all the associated momenta. Thus, upon Hermitian conjugation of $\mathcal O$, the fermionic field order in each $\mathcal O_i$ is reversed and each field object in $\mathcal O_i$ is conjugated according to its symbolic rule. For a linear combination, the coefficients $c_i$ are complex conjugated and each basis operator $\mathcal O_i$ is Hermitian conjugated following the rule for a product operator. 

\section{Quark propagation and Wick contractions}
\label{sec:wick}

Following the user-input of interpolator basis, symbolic representation of the correlation matrix to be evaluated is built, then all allowed quark propagations are identified, thereafter the correlation matrix is reformed in terms of quark lines or quark propagators. To this end, each correlation matrix element is expanded into a weighted sum of basis correlators, each built from single-term (basis) operators. This expanded basis aids in identifying redundancies and optimizing evaluation cost; the basis weights are multiplied into the coefficients of the corresponding Wick terms, so that every element of the original correlation matrix is carried through all later stages as a weighted sum of Wick terms and is summed at execution. The next step in \mypk\thinspace is in determining all allowed combinatorial pairing of quark-antiquark pairs for every quark flavor appearing in each of the correlation matrix element in the larger definite basis. For each quark flavor, the number of quark fields should be same as the number of antiquark fields in any given correlation matrix element considered. If this condition fails for any flavor, no contraction term would be evaluated for that matrix element. 

Once the quark-antiquark field balance is ensured for each flavor, every bijection between the quark and antiquark fields of each flavor is identified. $n_{\rm f}$ quark fields of a given flavor lead to $n_{\rm f}!$ possible pairings. Since the choices for different flavors are independent, the complete set of flavor-compatible pairings is given by $N_{\rm pair}=\prod_{\rm f} n_{\rm f}!$. Note that all quark-antiquark fields bijections generated are based on the flavor information explicitly specified in the elementary quark field objects that compose the hadronic operator. For example, $u$ and $d$ remain distinct quark flavors at this stage, even if isospin is assumed to be a good symmetry leading to the use of same light quark propagator/perambulator for either flavors at a later stage of the evaluation. Pairing of quark and antiquark fields of the same flavor when both occur on the same side of a two-point function or within the same composite insertion is also considered as a valid bijection. Retaining all allowed possibilities is necessary to realize the correlations, which may include closed quark loops or disconnected or mixed contraction topologies, to project the desired quantum numbers correctly. All admissible field pairings determined at this stage are retained explicitly and are passed on to the next stage discussed below. 

\begin{figure*}[pos=!tbp]
\centering
\begin{tikzpicture}[
    hadron/.style={ellipse, draw, fill=black!12, minimum width=0.34cm, minimum height=0.62cm, inner sep=0pt},
    baryon/.style={ellipse, draw, fill=black!12, minimum width=0.38cm, minimum height=0.95cm, inner sep=0pt},
    qline/.style={thick, postaction={decorate},
                  decoration={markings, mark=at position 0.55 with {\arrow{Latex[length=1.7mm]}}}},
    fl/.style={font=\footnotesize},
    tl/.style={font=\footnotesize, text=black!70},
    pcap/.style={font=\footnotesize, align=center, text width=3.9cm},
  ]
  \begin{scope}[shift={(0,0)}]
    \node[hadron] (s) at (0,0) {}; \node[hadron] (k) at (2.6,0) {};
    \draw[qline] (0.12,0.15) to[bend left=28] node[fl, above] {$u$} (2.48,0.15);
    \draw[qline] (2.48,-0.15) to[bend left=28] node[fl, below] {$d$} (0.12,-0.15);
    \node[tl] at (0,-1.05) {$t_0$}; \node[tl] at (2.6,-1.05) {$t$};
    \node[pcap] at (1.3,-1.6) {(a) meson, $\pi^+$};
  \end{scope}
  \begin{scope}[shift={(4.4,0)}]
    \node[hadron] (s) at (0,0) {}; \node[hadron] (k) at (2.6,0) {};
    \draw[qline] (s.east) .. controls +(0.95,0.8) and +(0.95,-0.8) .. (s.east);
    \draw[qline] (k.west) .. controls +(-0.95,-0.8) and +(-0.95,0.8) .. (k.west);
    \node[fl] at (0.55,0.45) {$u/d$}; \node[fl] at (2.05,0.45) {$u/d$};
    \node[tl] at (0,-1.05) {$t_0$}; \node[tl] at (2.6,-1.05) {$t$};
    \node[pcap] at (1.3,-1.6) {(b) closed loops, $\bar uu+\bar dd$};
  \end{scope}
  \begin{scope}[shift={(8.8,0)}]
    \node[baryon] (s) at (0,0) {}; \node[baryon] (k) at (2.6,0) {};
    \draw[qline] (0.17,0.27) -- (2.43,0.27);
    \draw[qline] (0.18,0) -- (2.42,0);
    \draw[qline] (0.17,-0.27) -- (2.43,-0.27);
    \node[fl, right] at (2.75,0.27) {$d$}; \node[fl, right] at (2.75,0) {$u$}; \node[fl, right] at (2.75,-0.27) {$u$};
    \node[tl] at (0,-1.05) {$t_0$}; \node[tl] at (2.6,-1.05) {$t$};
    \node[pcap] at (1.3,-1.6) {(c) baryon, direct};
  \end{scope}
  \begin{scope}[shift={(13.2,0)}]
    \node[baryon] (s) at (0,0) {}; \node[baryon] (k) at (2.6,0) {};
    \draw[qline] (0.17,0.27) -- (2.43,0.27);
    \draw[qline] (0.18,0) .. controls (1.2,0) and (1.4,-0.27) .. (2.43,-0.27);
    \draw[qline] (0.17,-0.27) .. controls (1.2,-0.27) and (1.4,0) .. (2.42,0);
    \node[fl, right] at (2.75,0.27) {$d$}; \node[fl, right] at (2.75,0) {$u$}; \node[fl, right] at (2.75,-0.27) {$u$};
    \node[tl] at (0,-1.05) {$t_0$}; \node[tl] at (2.6,-1.05) {$t$};
    \node[pcap] at (1.3,-1.6) {(d) baryon, exchange};
  \end{scope}

  \begin{scope}[shift={(0,-3.6)}]
    \node[hadron] (s1) at (0,0.6) {}; \node[hadron] (k1) at (2.6,0.6) {};
    \node[hadron] (s2) at (0,-0.6) {}; \node[hadron] (k2) at (2.6,-0.6) {};
    \draw[qline] (0.12,0.72) .. controls (1.1,0.8) and (1.5,-0.5) .. (2.48,-0.48);
    \draw[qline] (0.12,-0.72) .. controls (1.1,-0.8) and (1.5,0.5) .. (2.48,0.48);
    \draw[qline] (2.48,0.72) to[bend right=12] (0.12,0.48);
    \draw[qline] (2.48,-0.72) to[bend left=12] (0.12,-0.48);
    \node[fl] at (3.0,0.6) {$\pi^+$}; \node[fl] at (3.0,-0.6) {$\pi^+$};
    \node[tl] at (0,-1.25) {$t_0$}; \node[tl] at (2.6,-1.25) {$t$};
    \node[pcap] at (1.3,-1.8) {(e) two mesons, crossed $u$ lines};
  \end{scope}
  \begin{scope}[shift={(4.4,-3.6)}]
    \node[hadron] (s1) at (0,0.6) {}; \node[hadron] (k1) at (2.6,0.6) {};
    \node[hadron] (s2) at (0,-0.6) {}; \node[hadron] (k2) at (2.6,-0.6) {};
    \draw[qline] (0.12,0.6) -- (2.48,0.6);
    \draw[qline] (2.48,-0.6) -- (0.12,-0.6);
    \draw[qline] (k1.south) to[bend left=35] (k2.north);
    \draw[qline] (s2.north) to[bend left=35] (s1.south);
    \node[fl] at (3.0,0.6) {$\pi^+$}; \node[fl] at (3.0,-0.6) {$\pi^-$};
    \node[tl] at (0,-1.25) {$t_0$}; \node[tl] at (2.6,-1.25) {$t$};
    \node[pcap] at (1.3,-1.8) {(f) two mesons, equal-time lines};
  \end{scope}
  \begin{scope}[shift={(8.8,-3.6)}]
    \node[baryon] (s1) at (0,0.55) {}; \node[baryon] (k1) at (2.6,0.55) {};
    \node[hadron] (s2) at (0,-0.75) {}; \node[hadron] (k2) at (2.6,-0.75) {};
    \draw[qline] (0.17,0.75) -- (2.43,0.75);
    \draw[qline] (0.18,0.45) -- (2.42,0.45);
    \draw[qline] (0.12,-0.75) -- (2.48,-0.75);
    \draw[qline] (k2.north) to[bend right=30] (k1.south);
    \draw[qline] (s1.south) to[bend right=30] (s2.north);
    \node[fl] at (3.05,0.55) {$p$}; \node[fl] at (3.05,-0.75) {$\pi^+$};
    \node[tl] at (0,-1.25) {$t_0$}; \node[tl] at (2.6,-1.25) {$t$};
    \node[pcap] at (1.3,-1.8) {(g) baryon--meson, equal-time lines};
  \end{scope}
  \begin{scope}[shift={(13.2,-3.6)}]
    \node[baryon] (s1) at (0,0.55) {}; \node[baryon] (k1) at (2.6,0.55) {};
    \node[hadron] (s2) at (0,-0.75) {};
    \draw[qline] (0.17,0.82) -- (2.43,0.82);
    \draw[qline] (0.18,0.55) -- (2.42,0.55);
    \draw[qline] (0.17,0.28) -- (2.43,0.28);
    \draw[qline] (s2.east) .. controls +(0.95,0.8) and +(0.95,-0.8) .. (s2.east);
    \node[fl] at (3.05,0.55) {$p$}; \node[fl] at (-0.45,-0.75) {$\pi^0$};
    \node[fl] at (1.2,-0.75) {$u/d$};
    \node[tl] at (0,-1.25) {$t_0$}; \node[tl] at (2.6,-1.25) {$t$};
    \node[pcap] at (1.3,-1.8) {(h) $N\pi\to N$, pion loop at $t_0$};
  \end{scope}
\end{tikzpicture}
\caption{Representative quark-flow topologies, or Wick contraction diagrams, generated by \mypk. Ellipses are hadron operators with definite momentum at the source time $t_0$ (left) and the sink time $t$ (right); arrows follow quark propagation. (a) Single meson. (b) Disconnected Wick contraction diagram in the light isoscalar meson operator $(\bar uu+\bar dd)/\sqrt2$, with a $u$ or $d$ quark loop at the source and at the sink timeslice. (c, d) Direct and exchange Wick contractions of the proton. (e) Two mesons with crossed lines. (f) Two mesons with one equal-time quark propagation at each end. (g) Baryon and meson with equal-time quark propagation. (h) Nucleon--pion operator at the source and nucleon at the sink: the nucleon propagates from $t_0$ to $t$, while the pion annihilates into a closed quark loop at $t_0$. Table~\ref{tab:topology_counts} gives further details.}
\label{fig:topologies}
\end{figure*}
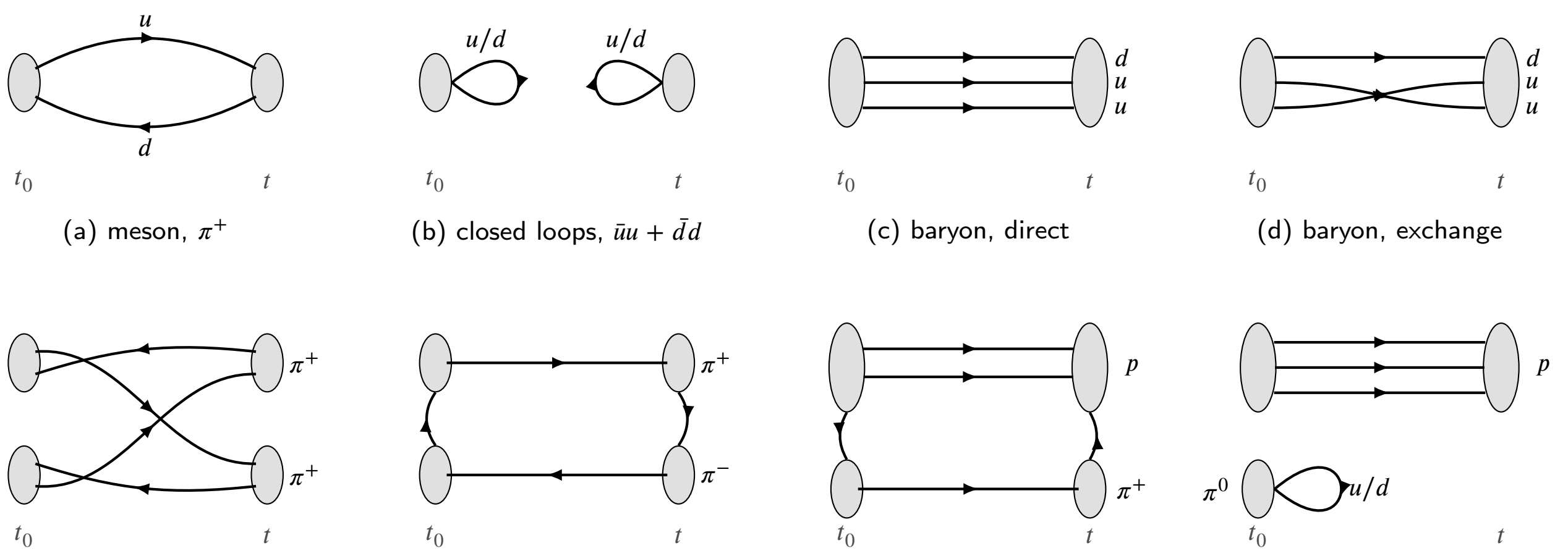

Complete tensor connection, within each admissible field pairings, leads to a diagram referred to as a Wick contraction diagram. Each Wick diagram is associated with a unique relative sign $s_{\alpha}$ and an algebraic coefficient $c_{\alpha}$, where $\alpha$ labels the Wick contraction as in Eq.~\ref{eq:distillation_contraction_schematic}. Representative Wick contractions are shown in Fig.~\ref{fig:topologies}, and Table~\ref{tab:topology_counts} lists the number of Wick contractions that \mypk\thinspace generates for various representative operators, separated into terms whose quark lines all connect source and sink, connected terms that contain equal-time quark lines, and disconnected terms with closed loops. The relative signs and the subsequent tensor-network construction follow from the same rules for all of them; closed quark loops are evaluated within the distillation subspace from ordinary all-time perambulators (Sec.~\ref{sec:distillation_lowering}).

\begin{table}[tbp]
\centering
\caption{Number of Wick terms generated by \mypk\thinspace for $C(t,t_0)=\langle\mathcal O(t)\,\mathcal O^\dagger(t_0)\rangle$, separated into terms whose quark lines all connect source and sink (src--snk), connected terms containing equal-time lines, and disconnected terms with closed loops. The single hadron operators are $\pi^+=\bar d\gamma_5 u$, $\bar u u$, $p=\epsilon_{abc}(u_a^\top C\gamma_5 d_b)u_c$ and $\Delta^{++}=\epsilon_{abc}(u_a^\top C\gamma_i u_b)u_c$. All two hadron operators ($\pi\pi$, $p\pi$, $pp$) are built from products of separately momentum projected single hadron operators. }
\label{tab:topology_counts}
\small
\setlength{\tabcolsep}{5pt}
\begin{tabular}{lrrrr}
\toprule
$\mathcal O$ & terms & src--snk & equal-time & loops \\
\midrule
$\pi^+$ & 1 & 1 & 0 & 0 \\
$\bar u u$ & 2 & 1 & 0 & 1 \\
$p$ & 2 & 2 & 0 & 0 \\
$\Delta^{++}$ & 6 & 6 & 0 & 0 \\
$\pi^+\pi^+$ & 4 & 4 & 0 & 0 \\
$\pi^+\pi^-$ & 4 & 1 & 2 & 1 \\
$p\pi^+$ & 12 & 6 & 6 & 0 \\
$pp$ ($I{=}1$, ${}^1S_0$) & 48 & 48 & 0 & 0 \\
\bottomrule
\end{tabular}
\end{table}

For a given Wick contraction, the relative sign $s_{\alpha}$ is fixed by the number parity of the fermionic field reordering required to arrive at the corresponding pairing expression from the original symbolic correlation matrix element. Other tensor objects such as spin matrices, color structures and spatial transformations within the correlation matrix element remain part of the correlator expression, and do not contribute to the fermion permutation signature. Consider a correlation matrix element with $N$ quark fields and thus $N$ antiquark fields, with the original user-defined fermionic field order $F$. For each given Wick contraction associated with the correlation matrix element, \mypk\thinspace builds the fermionic field ordering in which each quark is placed immediately left to its paired antiquark, for example
\begin{equation}
F_{\alpha} =(q_{i_1},\bar q_{j_1},\ldots, q_{i_N},\bar q_{j_N}).
\end{equation}
The fermion permutation sign is given by $s_{\alpha}=(-1)^{I_{\alpha}}$, where $I_{\alpha}$ is the number of pairwise exchanges of fields needed to bring $F$ into the order $F_{\alpha}$. 

We demonstrate this exercise with a simple example of correlation function with four fermion fields. Suppose the original fermionic order is $F=(q_1,\,q_2,\,\bar q_1,\,\bar q_2)$ with all four fields of the same flavor. For the pairing $[(q_1,\bar q_1), (q_2,\bar q_2)]$, the paired sequence is $F_{\alpha} =(q_{1},\bar q_{1},q_{2},\bar q_{2})$. Mapping $F$ to $F_{\alpha}$ requires one transposition and hence has sign $-1$. For the only other admissible pairing $[(q_1,\bar q_2), (q_2,\bar q_1)]$, the paired sequence is $F_{\alpha} =(q_{1},\bar q_{2},q_{2},\bar q_{1})$. In this case, mapping $F$ to $F_{\alpha}$ requires two transpositions and hence has sign $+1$. 

The algebraic coefficient $c_{\alpha}$ associated with each Wick contraction originates from the coefficients in the operator linear combinations within the user-defined basis. Each Wick term enters the logical correlator with the product $s_{\alpha}\,c_{\alpha}$. The sign and the algebraic coefficient are kept as separate fields of each Wick term and are multiplied only when the term is lowered to the logical correlator. This aids in making the origin of permutation sign traceable and thus
inspectable. 

\mypk\thinspace performs another backend modification to the user-supplied symbolic operator descriptions and the symbolic correlation matrix expressions, while retaining their original forms at the frontend. In this procedure, each symbolic description is associated with a structural identity that is independent of the names of summed (dummy) indices and position labels and of how chains of field operations are nested, while the order of factors and the open index labels remain part of the identity. While the user enjoys the freedom to define the same operator in visibly different forms, the backend structural identity aids in identifying them and in lifting any redundancy in their evaluation. This modification also involves expanding and simplifying nested field operations into an internal composite operation, while preserving fermion field ordering, spin and color structures, spatial transformations and momentum assignments. The backend structural identity aids in identifying repeated occurrences of the same normalized symbolic operators and correlator matrix element expressions. This facilitates a single time evaluation of identical expressions and reuse to avoid redundant symbolic work. There is a separate optimization program later in the \mypk\thinspace workflow, when exact global common-subexpression analysis can canonicalize reusable tensor-network components and combine exactly equal numerical objects within each correlator. 

By the end of this stage, \mypk\thinspace identifies all Wick contractions, the associated relative permutation sign, the algebraic weights, and the symbolic operator structures that accompany those propagation lines for every correlation matrix element to be evaluated.  For each Wick contraction in every correlation matrix element, this stage associates and outputs a symbolic description that is independent of lattice specifics and other numerical details involved in the actual computation, 
\begin{equation}
W_\alpha =\left( s_\alpha, \,c_\alpha,\, \mathcal P_\alpha,\, \mathcal E_\alpha,\, \mathcal C_\alpha\right).
\label{eq:wick_term_contents}
\end{equation}
where $s_\alpha$ is the fermionic sign and $c_\alpha$ is the algebraic coefficient. $\mathcal P_\alpha$ is the set of quark-antiquark field pairings. Each pairing carries information of the quark endpoint and the antiquark endpoint of one quark propagation line, including its flavor and symbolic spin, color, and position labels. Its endpoints are references to the field occurrences in the basis correlator, which together with that correlator associate field-level derivatives, transports, shifts, and spatial representations with the correct endpoints. $\mathcal E_\alpha$ is the contracted symbolic expression that retains all the non-fermionic tensor content, such as spin matrices, color tensors, gauge-link objects, their connectivity, etc., inherited from the original operators, in which bare contracted fields are replaced by symbolic propagators, while fields carrying derivatives, transports or shifts keep their operation wrappers. $\mathcal C_\alpha$ records the connected components of the position (locality) graph formed by the quark lines, together with the correlator insertions they touch; it provides a coarse connected/disconnected classification of the term. 

\section{Redefining in the distillation language}
\label{sec:distillation_lowering}
The output from Wick contraction stage is passed on to the next stage involving (symbolic) reduction of tensors from space-color representation to that of distillation vectors (see Eq.~\ref{eq:distill}). This consists of a few substages: substituting the space color indices of the field endpoints with the distillation vector indices, building symbolic objects representing operator dependent kernels/elementals and quark lines tensors, and constructing tensor-contracted form of correlation matrix elements in the distillation space. We discuss this procedure in this section. 

A field endpoint refers to a (anti-)quark field within any hadron operator and is conventionally associated with a spatial index coupled with any derivative, transport or shift operations. The field endpoints constitute the hadronic operators at the source or sink timeslice, for which the package supports conversion of the space-color index represented field endpoints to the distillation endpoints. It also supports retaining the spatial index for current insertions at intermediate timeslices, supporting the use of generalized perambulators and thus the construction of three-point correlation functions/matrices. For a distillation endpoint, the spatial-color explicit fermion field objects are projected to the distillation vector space following
\begin{equation}
q(\mathbf x, a) \longmapsto q(V),
\qquad
\bar q(\mathbf x, a) \longmapsto \bar q(V^\dagger),
\label{eq:laph_endpoint_realization}
\end{equation}
whereas an explicitly POINT field endpoint retains the space-color information. 

The hadronic operators at the source, sink and current insertion timeslices in the distillation vector space are constructed from the spatial-color components of the field endpoints. The code assigns a unique kernel identity to each spatial--color connected component of an operator (one per local hadron in a multi-hadron operator) that includes various information on the projected-field orientations, field-local operations, color tensors, explicit gauge connectivity, and momentum assignments in the original operator definition. Spin is deliberately excluded from this kernel identity; the flavor of each field is retained, because a data provider may supply flavor-dependent eigenvectors. Keeping the space-color representation separate from the spin-flavor structure reduces the cost and memory of the kernels: a local baryon kernel without spin has $N_v^3$ entries, whereas folding its three spin indices into it would give $64\,N_v^3$, and one spin-free kernel is shared by all spin structures of the same spatial construction and by every reordering of the same fields, while kernels whose fields carry different flavors remain distinct resources, which also exposes reuse across Wick contractions in the later optimization stages. The kernel recipe behind this identity carries the operator structure forward to the contraction stages, to determine the required numerical tensors. 

For example, a local bilinear representing simple meson, whose two distillation-projected quark fields are contracted to a color singlet giving a two-index kernel of the schematic form
\begin{equation}
M_{kl}^{(\mathbf p)}(t) =
\sum_{\mathbf x,a}
 e^{+i\mathbf p\cdot\mathbf x}
 \left[\mathcal D_{\bar q}V(t)\right]^*_{\mathbf x a k}
 \left[\mathcal D_q V(t)\right]_{\mathbf x a l},
\label{eq:bilinear_spatial_kernel}
\end{equation}
where \(\mathcal D_q\) and \(\mathcal D_{\bar q}\) stand for the spatial operations attached to the quark and antiquark field endpoints. The same construction applies without changing the lowering algorithm when the hadronic operator has a higher rank. For a local three-quark color-singlet structure, the spatial part has the schematic form
\begin{align}
B_{k_1k_2k_3}^{(\mathbf p)}(t)
&=
\sum_{\mathbf x,a,b,c}
 e^{+i\mathbf p\cdot\mathbf x}
 \epsilon_{abc}
 \left[\mathcal D_{q_1} V(t)\right]_{\mathbf x a k_1}\nonumber\\
&\qquad\times
 \left[\mathcal D_{q_2} V(t)\right]_{\mathbf x b k_2}
 \left[\mathcal D_{q_3} V(t)\right]_{\mathbf x c k_3}.
\label{eq:three_quark_spatial_kernel}
\end{align}
Note that spin does not enter the definition of the kernels. Flavor is part of their identity, but with the built-in eigenvector provider, which uses one eigenvector basis for all flavors, kernels that differ only in flavor have the same values. The difference between the bilinear and three-quark examples therefore appears only through the number of field endpoints and their explicit color connectivity. Another important aspect is that the momentum projection is part of the spatial resource itself. Each position locality group involved in a hadronic operator is associated with a definite momentum assignment and \mypk\thinspace retains this assignment explicitly using the phase convention $e^{+i\mathbf p\cdot\mathbf x}$ for each and the product of those phase factors. This keeps constituent momenta visible for bilocal and multi-local constructions rather than replacing them by only the total momentum.

The next objects to be redefined in the distillation language are the quark propagation lines that connect the field endpoint pairs, identified in the Wick contraction pairings in the previous stage. These objects in the distillation setup are referred to as the perambulators. An open path connecting two field endpoints with no current insertion is represented by an ordinary perambulator object defined in Eq.~\ref{eq:peram}. The package also accepts generalized perambulators, which involve an explicit current insertion and thus support the evaluation of three point functions/matrices. The perambulator objects are also associated with a unique propagator identity that carries the kind of propagation (ordinary or generalized), and its flavor content. The kernels and ports that the perambulator connects are recorded separately, as its placement in the tensor network. Since kernels and ports are not part of the propagator identity, every propagation line of the same kind and flavor content, in every Wick term and every correlation-matrix element, refers to one resource, and one compiled requirement serves every timeslice and configuration.

The generalized perambulator object retains substantially more information than an ordinary perambulator. In addition to the ordinary perambulator attributes, it carries information on the inserted current, i.e. its POINT-like spatial kernel, current momentum assignment, incoming and outgoing flavor, and the current insertion spin structures in the propagator identity. \mypk\thinspace does not compute generalized perambulators; they have to be supplied by the user, typically from sequential or stochastic inversions. The generalized perambulators are specified by a short user function named in the job file; for each one it receives its description (the flavors, $\Gamma$ and momentum of the current) and returns the file and dataset that hold the data. 

For example, consider the flavor-changing current
$J_\Gamma=\bar s\,\Gamma\,d$ between pseudoscalar interpolators forming a three point correlation function 
\begin{equation}
C_\Gamma(t,t_1,t_0)=\big\langle\, [\bar u\gamma_5 s]_{\mathbf p'}(t)\;
[\bar s\,\Gamma\,d]_{\mathbf q}(t_1)\; [\bar d\gamma_5 u]_{\mathbf p}(t_0)\,\big\rangle ,
\label{eq:threept_example}
\end{equation}
with a momentum on each of the three operators. Every operator is written as it enters the correlation function and carries the phase of its own momentum, so the three momenta are chosen to add up to zero, $\mathbf p'+\mathbf q+\mathbf p=0$. The evaluation request can be defined as 
\begin{verbatim}
def build(G=gammaX):
    snk = Operator(ubar("x", "a", "al")
          * gamma5("al", "be") * s("x", "a", "be"),
          momentum=(0, 0, 1))
    J = Operator(point(sbar("z", "b", "ga"))
          * G("ga", "de") * point(d("z", "b", "de")),
          momentum=(1, 0, -1))
    src = Operator(dbar("y", "c", "ep")
          * gamma5("ep", "ze") * u("y", "c", "ze"),
          momentum=(-1, 0, 0))
    return {"C3": Correlator(snk, J, src)}
\end{verbatim}
The flavor changes between $s$ and $d$ at the current, and \mypk\thinspace absorbs the current into one generalized perambulator, so that the correlation function can be written in the form
\begin{align}
C_\Gamma(t,t_1,t_0)={}&-\,\mathrm{Tr}\big[\Phi_{\gamma_5,\mathbf p'}(t)\;
\tau^{\Gamma,\mathbf q}_{d\to s}(t,t_1,t_0)\nonumber\\
&\qquad\times\Phi_{\gamma_5,\mathbf p}(t_0)\;\tau_u(t_0,t)\big],\\
\tau^{\Gamma,\mathbf q}_{d\to s}(t,t_1,t_0)={}&\sum_{\mathbf z}e^{+i\mathbf q\cdot\mathbf z}\,
V^\dagger(t)\,M_s^{-1}(t;\mathbf z,t_1)\nonumber\\
&\qquad\times\Gamma\,M_d^{-1}(\mathbf z,t_1;t_0)\,V(t_0),
\label{eq:generalized_peram}
\end{align}
where $\Phi$ are the kernel functions defining the source and sink interpolators, with their distillation and spin indices open. The current's $\Gamma$ and $\mathbf q$ do not appear as separate tensors; they are part of the identity of the generalized perambulator. 

The standard perambulator expected by Wick contraction object is ordered as sink vector and sink spin followed by source vector and source spin. When the data contain only fixed-source perambulators $\tau_f(t,t_0)$, a propagation in the reverse direction is reconstructed with $\gamma_5$-Hermiticity, $\tau_f(t_0,t)=\gamma_5\,\tau_f(t,t_0)^\dagger\gamma_5$, whereas perambulators stored for all source times are read directly in both directions. The Hermitian adjoint exchanges both vector and spin indices at the source as well as sink timeslices. The storage conventions of the external files are declared in the job and applied once, when each perambulator is read. Following this, the axes are permuted to the canonical order and a $q^\dagger$ source-field convention is converted to $\bar q$. Closed fermion cycles, which include the connected meson trace as well as self-annihilation loops at any operator time, use a related lowering: each Wick segment of the cycle is an ordinary perambulator and the operator transitions supply the kernels and spin matrices that close the trace. Self-annihilation quark loops are built with equal-time perambulator $\tau_f(t,t)$ at the time of the operator by self-contracting and tracing $\tau_f(t,t)$ with the appropriate spatial and spin structure sandwiched within the trace. A POINT-like quark loop, as produced by a flavor-diagonal current insertion in a three-point function, cannot be built from perambulators, which do not contain the point-to-all propagator. It requires stochastic, all-to-all or precomputed-loop information and is rejected by the present exact-distillation lowering. POINT fields are therefore supported only inside generalized perambulators.

In the next and final substage of lowering procedure, \mypk\thinspace converts each identified complete Wick contraction into a primitive tensor-contraction graph, from the distillation objects constructed above. This graph is the first representation in which the numerical contraction problem is explicit. It contains the tensor objects (the operator kernels, perambulators, and spin structures) that participate in the Wick contraction terms and the indices that connect them. Each graph is validated before it is passed to the subsequent stages; its connected components receive canonical identities in the next stage. Node identifiers are ensured to be unique, an index carried by two tensor legs is summed while an index on a single leg remains an open output index (for example an open spin index), and indices with more than two instances are ruled out. These checks establish the well-formed numerical tensor network independent of any subsequent contraction path. 

Before we leave this stage, note that none of the distillation objects built are based on any concrete numerical data. No files, no information on source/sink timeslices, tensor shape, physical buffer, CPU/GPU placement, or contraction workflow has been fed in or sought out. The output from this distillation lowering stage builds explicit tensor components to be evaluated, while remaining independent of any lattice specifics, any particular data layout or machine.

\section{Execution and Optimization Framework}
\label{sec:execution}

The next stage in the procedure concerns building the plan and finding optimizations for a cost-efficient evaluation. As a first step to this, \mypk\thinspace builds a directed acyclic graph (DAG) representation detailing the workflow in evaluating all the tensor contraction graphs of the requested correlation matrix, built in the previous stage. This step proceeds through the identification of primitive resource requirements from their exact resource identities. Then the connectivity in each tensor contraction graphs is canonicalized independently of any proxy node names and dummy contraction-index labels. This canonicalization aids in identifying similarity across Wick terms as well as correlation matrix elements and thus identifying reusable resources as well as redundant evaluations. This DAG representation is then progressively refined through global exact reuse, contraction-order selection, partial-intermediate reuse to arrive at the plan, which lists the contraction steps of all graphs in the order in which they are executed: each step follows the steps that produce its inputs, and among the steps that are ready, the one that frees the most memory is taken first. A step contracts two tensors into one, so a connected network of $n$ tensors is evaluated in $n-1$ steps; a step shared by several networks is performed only once. Every step records its input values and interfaces, the indices eliminated at that step, the outputs, an estimated operation count, and the size of the resulting intermediate objects. 

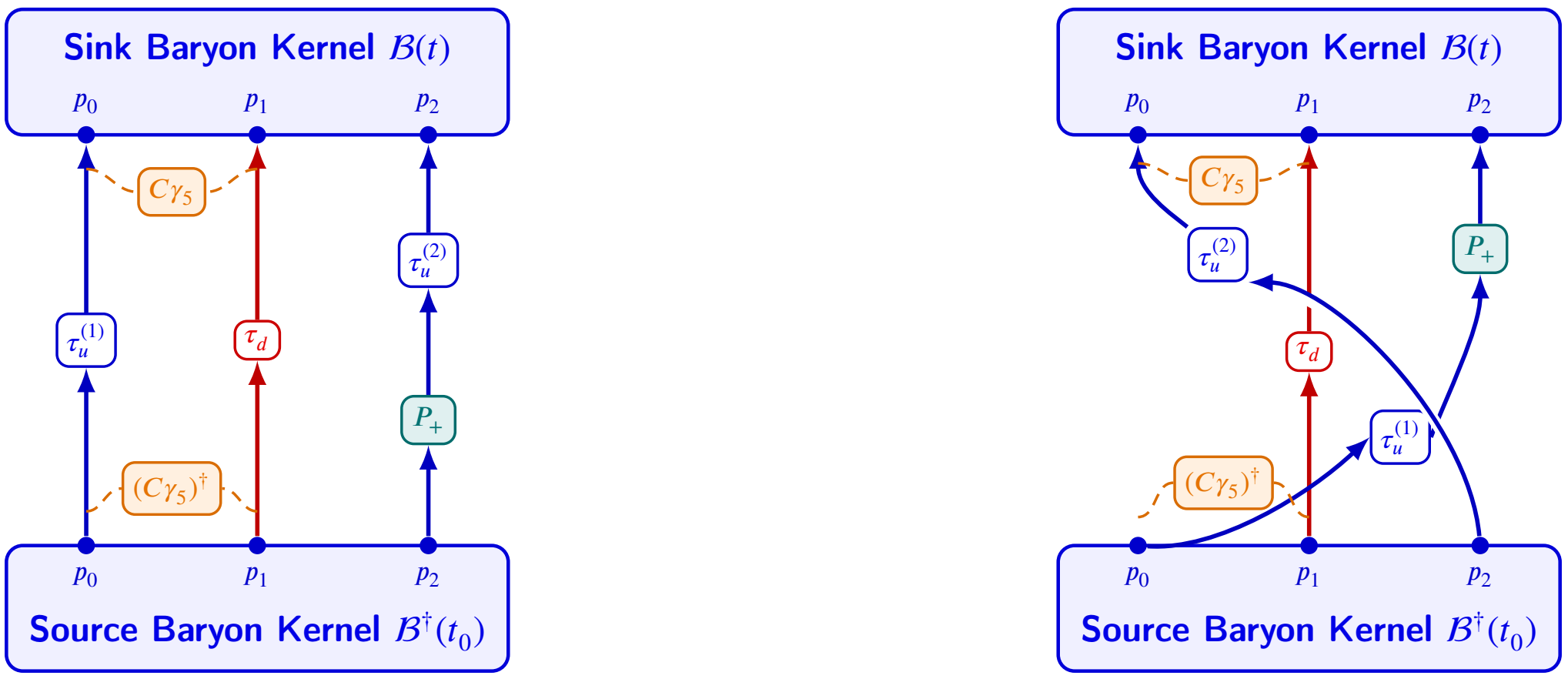
\begin{figure*}[pos=!b]
\centering
\begin{subfigure}[b]{0.48\textwidth}
\centering
\begin{tikzpicture}[scale=0.98, >=latex]
  \tikzset{
    kernel/.style={draw=blue!85!black, fill=blue!6, rounded corners=4pt, line width=0.85pt},
    kernel_title/.style={font=\small\bfseries, text=blue!90!black},
    port_dot/.style={circle, fill=blue!80!black, inner sep=1.5pt},
    port_lbl/.style={font=\scriptsize\bfseries, text=blue!80!black},
    diquark/.style={draw=orange!85!black, fill=orange!12, rounded corners=3pt, font=\scriptsize\bfseries, text=orange!90!black, inner sep=2.5pt, line width=0.65pt},
    proj/.style={draw=teal!85!black, fill=teal!12, rounded corners=3pt, font=\scriptsize\bfseries, text=teal!90!black, inner sep=2.5pt, line width=0.65pt},
    prop_u/.style={draw=blue!80!black, fill=white, rounded corners=3pt, font=\scriptsize\bfseries, text=blue!90!black, inner sep=2.2pt, line width=0.65pt},
    prop_d/.style={draw=red!80!black, fill=white, rounded corners=3pt, font=\scriptsize\bfseries, text=red!90!black, inner sep=2.2pt, line width=0.65pt},
    fermi_u/.style={->, line width=1.05pt, draw=blue!75!black},
    fermi_d/.style={->, line width=1.05pt, draw=red!75!black},
    spin_link/.style={densely dashed, draw=orange!85!black, line width=0.65pt}
  }

  \draw[kernel] (-2.2, 3.5) rectangle (2.2, 4.6);
  \node[kernel_title] at (0, 4.25) {Sink Baryon Kernel $\mathcal{B}(t)$};
  \node[port_dot] (snk_p0) at (-1.5, 3.5) {};
  \node[port_dot] (snk_p1) at (0.0, 3.5) {};
  \node[port_dot] (snk_p2) at (1.5, 3.5) {};
  \node[port_lbl, above=2pt] at (snk_p0) {$p_0$};
  \node[port_lbl, above=2pt] at (snk_p1) {$p_1$};
  \node[port_lbl, above=2pt] at (snk_p2) {$p_2$};

  \draw[kernel] (-2.2, -1.2) rectangle (2.2, -0.1);
  \node[kernel_title] at (0, -0.85) {Source Baryon Kernel $\mathcal{B}^\dagger(t_0)$};
  \node[port_dot] (src_p0) at (-1.5, -0.1) {};
  \node[port_dot] (src_p1) at (0.0, -0.1) {};
  \node[port_dot] (src_p2) at (1.5, -0.1) {};
  \node[port_lbl, below=2pt] at (src_p0) {$p_0$};
  \node[port_lbl, below=2pt] at (src_p1) {$p_1$};
  \node[port_lbl, below=2pt] at (src_p2) {$p_2$};

  \node[diquark] (g_snk) at (-0.75, 3.0) {$C\gamma_5$};
  \node[diquark] (g_src) at (-0.75, 0.4) {$(C\gamma_5)^\dagger$};
  \node[proj]    (proj)  at (1.5, 1.0)  {$P_+$};

  \node[prop_u] (tau_u1) at (-1.5, 1.7) {$\tau_u^{(1)}$};
  \node[prop_d] (tau_d)  at (0.0, 1.7)  {$\tau_d$};
  \node[prop_u] (tau_u2) at (1.5, 2.4)  {$\tau_u^{(2)}$};

  \draw[fermi_u] (src_p0) -- (tau_u1);
  \draw[fermi_u] (tau_u1) -- (snk_p0);

  \draw[fermi_d] (src_p1) -- (tau_d);
  \draw[fermi_d] (tau_d) -- (snk_p1);

  \draw[fermi_u] (src_p2) -- (proj);
  \draw[fermi_u] (proj) -- (tau_u2);
  \draw[fermi_u] (tau_u2) -- (snk_p2);

  \draw[spin_link] (-1.5, 3.2) to[out=0, in=180] (g_snk.west);
  \draw[spin_link] (0.0, 3.2)  to[out=180, in=0] (g_snk.east);
  \draw[spin_link] (-1.5, 0.2) to[out=0, in=180] (g_src.west);
  \draw[spin_link] (0.0, 0.2)  to[out=180, in=0] (g_src.east);

\end{tikzpicture}
\caption{Direct contraction graph (sign \(+1\))}
\label{fig:proton_graph_direct}
\end{subfigure}
\hfill
\begin{subfigure}[b]{0.48\textwidth}
\centering
\begin{tikzpicture}[scale=0.98, >=latex]
  \tikzset{
    kernel/.style={draw=blue!85!black, fill=blue!6, rounded corners=4pt, line width=0.85pt},
    kernel_title/.style={font=\small\bfseries, text=blue!90!black},
    port_dot/.style={circle, fill=blue!80!black, inner sep=1.5pt},
    port_lbl/.style={font=\scriptsize\bfseries, text=blue!80!black},
    diquark/.style={draw=orange!85!black, fill=orange!12, rounded corners=3pt, font=\scriptsize\bfseries, text=orange!90!black, inner sep=2.5pt, line width=0.65pt},
    proj/.style={draw=teal!85!black, fill=teal!12, rounded corners=3pt, font=\scriptsize\bfseries, text=teal!90!black, inner sep=2.5pt, line width=0.65pt},
    prop_u/.style={draw=blue!80!black, fill=white, rounded corners=3pt, font=\scriptsize\bfseries, text=blue!90!black, inner sep=2.2pt, line width=0.65pt},
    prop_d/.style={draw=red!80!black, fill=white, rounded corners=3pt, font=\scriptsize\bfseries, text=red!90!black, inner sep=2.2pt, line width=0.65pt},
    fermi_u/.style={->, line width=1.05pt, draw=blue!75!black},
    fermi_d/.style={->, line width=1.05pt, draw=red!75!black},
    spin_link/.style={densely dashed, draw=orange!85!black, line width=0.65pt}
  }

  \draw[kernel] (-2.2, 3.5) rectangle (2.2, 4.6);
  \node[kernel_title] at (0, 4.25) {Sink Baryon Kernel $\mathcal{B}(t)$};
  \node[port_dot] (x_snk_p0) at (-1.5, 3.5) {};
  \node[port_dot] (x_snk_p1) at (0.0, 3.5) {};
  \node[port_dot] (x_snk_p2) at (1.5, 3.5) {};
  \node[port_lbl, above=2pt] at (x_snk_p0) {$p_0$};
  \node[port_lbl, above=2pt] at (x_snk_p1) {$p_1$};
  \node[port_lbl, above=2pt] at (x_snk_p2) {$p_2$};

  \draw[kernel] (-2.2, -1.2) rectangle (2.2, -0.1);
  \node[kernel_title] at (0, -0.85) {Source Baryon Kernel $\mathcal{B}^\dagger(t_0)$};
  \node[port_dot] (x_src_p0) at (-1.5, -0.1) {};
  \node[port_dot] (x_src_p1) at (0.0, -0.1) {};
  \node[port_dot] (x_src_p2) at (1.5, -0.1) {};
  \node[port_lbl, below=2pt] at (x_src_p0) {$p_0$};
  \node[port_lbl, below=2pt] at (x_src_p1) {$p_1$};
  \node[port_lbl, below=2pt] at (x_src_p2) {$p_2$};

  \node[diquark] (x_g_snk) at (-0.75, 3.1) {$C\gamma_5$};
  \node[diquark] (x_g_src) at (-0.75, 0.45) {$(C\gamma_5)^\dagger$};
  \node[proj]    (x_proj)  at (1.5, 2.5)   {$P_+$};

  \node[prop_d] (x_tau_d)  at (0.0, 1.6)   {$\tau_d$};
  \node[prop_u] (x_tau_u1) at (0.8, 0.85)  {$\tau_u^{(1)}$};
  \node[prop_u] (x_tau_u2) at (-0.8, 2.45) {$\tau_u^{(2)}$};

  \draw[fermi_d] (x_src_p1) -- (x_tau_d);
  \draw[fermi_d] (x_tau_d) -- (x_snk_p1);

  \draw[fermi_u] (x_src_p0) .. controls (-0.8, -0.15) and (0.0, 0.35) .. (x_tau_u1.west);
  \draw[fermi_u] (x_tau_u1.east) .. controls (1.2, 1.2) and (1.5, 1.8) .. (x_proj.south);
  \draw[fermi_u] (x_proj.north) -- (x_snk_p2);

  \draw[line width=2.8pt, white] (x_src_p2) .. controls (1.4, 1.0) and (0.2, 2.2) .. (x_tau_u2.south east);
  \draw[fermi_u] (x_src_p2) .. controls (1.4, 1.0) and (0.2, 2.2) .. (x_tau_u2.south east);

  \draw[line width=2.8pt, white] (x_tau_u2.north west) .. controls (-1.2, 2.8) and (-1.5, 3.0) .. (x_snk_p0);
  \draw[fermi_u] (x_tau_u2.north west) .. controls (-1.2, 2.8) and (-1.5, 3.0) .. (x_snk_p0);

  \draw[spin_link] (-1.5, 3.25) to[out=0, in=180] (x_g_snk.west);
  \draw[spin_link] (0.0, 3.25)  to[out=180, in=0] (x_g_snk.east);
  \draw[spin_link] (-1.5, 0.15) to[out=0, in=180] (x_g_src.west);
  \draw[spin_link] (0.0, 0.15)  to[out=180, in=0] (x_g_src.east);

\end{tikzpicture}
\caption{Exchange contraction graph (sign \(-1\))}
\label{fig:proton_graph_exchange}
\end{subfigure}
\caption{\mypk\thinspace representation of two Wick contraction topologies in proton correlators, following the convention of distillation hadron contraction networks~\cite{10.1145/3592979.3593409}. Blue and red solid directed lines represent up- and down-quark propagations ($\tau_u,\,\tau_d$) connecting source to sink indices. Dashed orange links represent Dirac spinor contractions with the diquark matrix $C\gamma_5$ and the positive-parity projector $P_+$.}
\label{fig:proton_graphs_comparison}
\end{figure*}

\mypk\thinspace further utilizes \texttt{opt\_einsum}~\cite{Smith:2018opteinsum} to choose the efficient algebraic path for each unique connected component, which is preserved in the backend. Searches for the efficient algebraic path are assessed through the number of distillation modes $N_v$, which is treated as an optimization parameter, rather than part of the graph identity. This means that a change in $N_v$ can therefore change the algebraic contraction path without affecting the connectivity of the graphs. The chosen algebraic contraction path identifies and records recursively defined inputs, reusable primitive tensors and their exact logical connections. This facilitates in further identifying similarities and commonalities across different structurally identical (same inputs, placements and index connectivity) partial as well as full connections, even when they arise or belong to different graphs. The final DAG representation of the graph carries a deduplicated set of tensor objects and their connectivity information necessary to build final correlation matrix. The four classes of tensor objects preserved in the DAG representation include the primitive tensors (perambulators, operator and spin structures), reusable intermediate tensor objects, the Wick terms (products of the component graphs with their coefficients) and the final correlation matrix elements. In the following, we discuss the execution and optimization programs in this stage of \mypk\thinspace workflow, taking example of nucleon correlation function evaluation. 

The conventional positive-parity proton interpolating operator in the rest frame is given by
\begin{equation}
\mathcal O_N(t) = \epsilon_{abc} \left[ u^a(t)^\top C\gamma_5 d^b(t) \right] P_+ u^c(t),
\label{eq:proton_interpolator_def}
\end{equation}
where $C = \gamma_2\gamma_4$ is the charge-conjugation matrix, $\epsilon_{abc}$ is the totally antisymmetric Levi-Civita tensor representing the baryon color structure, and $P_+ = \frac{1}{2}(\mathbb I + \gamma_4)$ is the positive-parity projector. Note that all quark fields in this operator correspond to the same position locality group. The quark fields contained in the $[...]$ forms the diquark, while the third one outside is referred to as the passenger quark, which carries the open Dirac index of the baryon. The corresponding Euclidean two-point correlation function is
\begin{equation}
C_{NN}(t, t_0) = \left\langle \mathcal O_N(t) \, \overline{\mathcal O}_N(t_0) \right\rangle,
\label{eq:proton_correlator_def}
\end{equation}
where $\overline{\mathcal O}_N(t_0)$ is the adjoint proton operator defined at the source timeslice. $C_{NN}$ carries the two open Dirac indices of the proton operators at source and at the sink and thus,  $C_{NN}$ is a $4\times4$ matrix for any fixed value of $t$ and $t_0$. 

There are two distinct Wick contractions for the proton two-point correlator as demonstrated in Fig.~\ref{fig:proton_graphs_comparison}. The left diagram is the direct term, for which the fermionic permutation sign is $s_{\alpha} = +1$. Here, the two quark fields forming the diquark at the sink contract with their respective counterparts at the source, and the passenger $u$ quark likewise propagates directly from source to sink. The right diagram is the exchange term, with $s_\alpha = -1$. In this case, the $\bar u$ in the source diquark contracts with the passenger $u$ at the sink, while the passenger $\bar u$ at the source contracts with the $u$ in the sink diquark. For a given source $t_0$ and sink $t$ timeslice, the primitive tensors involved in this computation are: the baryon kernels at $t$ and $t_0$ each carrying three indices running over the distillation modes; the quark perambulators for different flavors ($u$ and $d$ for proton) each carrying four indices corresponding to sink distillation mode, sink spin, source distillation mode and source spin; and spin structures $C\gamma_5$ and $P_+$. Even when the perambulators for both flavors are read from the same perambulator file, \mypk\thinspace keeps them as separate resources by default (Sec.~\ref{sec:execution}). 

To facilitate and streamline the optimization program, \mypk\thinspace maintains four distinct object identifiers. 
\begin{enumerate}
    \item \textbf{\texttt{correlator\_id}}: Assigned to represent the physical correlator evaluated (e.g., \texttt{c\_proton} or \texttt{C\_0\_1}). 
    \item \textbf{\texttt{node\_id}}: Distinctly assigned to each primitive tensors necessary to build the graph. (e.g., \texttt{kernel\_0}, \texttt{propagation\_1}, \texttt{spin\_2}).
    \item \textbf{\texttt{resource\_id}}: A globally unique identifier for the underlying physical/numerical data source, which can be perambulators or pre-built operator kernels. The direct and exchange graphs have distinct connectivity but share the same resources and thus the \texttt{resource\_id} values.
    \item \textbf{\texttt{component\_id}}: Identifies each unique, canonicalized, connected tensor contraction graph. This identifier is invariant under the renaming of dummy indices and the reordering of tensors; open indices, such as the two Dirac indices of the proton, remain part of the graph. When a single Wick term has several disconnected parts, as in a nucleon--pion term in which the pion propagates separately from the nucleon, each part receives its own identifier.
\end{enumerate}
All these identifiers are unique for the respective objects they represent and hence are invariant while executing the program across different configurations from the same ensemble. 

Different graphs could be isomorphic or can have similarities, and thus share same mathematical (sub)expressions and can have similar final or intermediate contraction results, although their non-canonicalized representations are distinct. This can be either due to dummy index namings or due to arbitrariness in the tensor index and tensor connectivity ordering. To recognize these identities/similarities, the \texttt{Redstar} \cite{10.1145/3592979.3593409} package uses the general-purpose package \texttt{nauty}~\cite{McKay:2014}, which performs \emph{canonical (re)labeling} of all the graph indices and connectivity. In the present realization, \mypk\thinspace focuses on systems with up to nine quarks, whose graphs contain a few tens of primitive tensors, so that an exhaustive starting node/tensor search is affordable. If the scope of the investigation goes beyond nine-quark operators, one may resort to more involved canonical labeling algorithms realized in general-purpose package \texttt{nauty}. Note that the process of \emph{canonical (re)labeling} is a one-time overhead and is invariant across lattice QCD ensemble specifics as well as across configurations. \mypk\thinspace therefore uses a lightweight, self-contained canonicalization strategy based on three intuitive steps discussed below.
\begin{itemize}
    \item \textit{Traversal and index standardization}: Starting from a tensor, the graph is walked through, following the fixed index ordering in each tensor. The tensor indices are renamed in order of encounter: $i_0, i_1, \ldots$ for contracted indices and $o_0, o_1, \ldots$ for open indices.
    \item \textit{Unbiased root selection}: A closed loop of contracted tensors has no natural first tensor, so the walk is started from every tensor in turn. Each start yields a candidate string.
    \item \textit{Deterministic tie-breaking}: The lexicographically smallest candidate string is selected as the canonical form.
\end{itemize}

Since this procedure exhaustively considers all starting tensors and deterministically selects the lexicographically minimal description, the resulting graph representation serves as its canonical form. \mypk\thinspace generates a cryptographic fingerprint, the \texttt{component\_id}, for each connected tensor contraction graph in its canonical form. Two contraction graphs receive the same \texttt{component\_id} exactly when they are identical up to the order of their tensors and the names of their indices, with the same tensor kinds, resources, operator-slot placements and index connections. When two fingerprints agree, the full canonical strings are compared for graph equivalence. If two graphs share the same \texttt{component\_id}, the tensor contraction is scheduled and computed only once, and the result is reused wherever needed in the global correlation matrix. Additionally, within a single correlator sum, if two Wick terms have identical components but coefficients that sum to zero, \mypk\thinspace removes the terms from the plan, so that no operations or buffer memory are wasted. Equalities that rely on tensor symmetries or on physics relations such as Hermiticity are not detected by this structural test; for example, the off-diagonal elements $C_{12}$ and $C_{21}$ of a nucleon correlation matrix built from $C\gamma_5$ and $C\gamma_4\gamma_5$ diquarks carry different spin matrices at the source and the sink and are computed separately.

Many components or graphs of a multi-hadron correlator could be equal for other reasons: a local baryon kernel is antisymmetric in its three mode indices, a local multi-baryon operator may have intrinsic symmetries under quark field exchanges, fixed spin matrices obey identities such as $(C\gamma_\mu)^\top=C\gamma_\mu$, and, when isospin symmetry $m_u=m_d$ is intact, terms that differ only by exchanging the two flavors coincide. Such equalities can be determined by evaluating the correlators using dummy random quark propagators, as done for the determinants of nuclear correlation functions in Ref.~\cite{PhysRevD.110.114505}. We applied this test to four correlation functions of operators built from hadrons at rest, defined in Table~\ref{tab:wick_identities}: a nucleon--pion operator with $I_3=+\tfrac12$, two-nucleon operators in the ${}^1S_0$ and ${}^3S_1$ channels, and a three-nucleon operator $p\,[nn]$. The numerically evaluated Wick terms were grouped into proportionality classes based on eight random quark propagator samples. Table~\ref{tab:wick_identities} lists the results for correlation function without ($m_u\ne m_d$) and with ($m_u=m_d$) isospin symmetry. With as many as eight random samples, the probability of grouping two inequivalent Wick terms is negligible for all four systems.

\begin{table}[t]
\centering
\caption{Random-matrix test of complete Wick terms: number of Wick terms and number of proportionality classes that do not cancel, for distinct ($u\neq d$) and identical ($u=d$) $u$ and $d$ perambulators, at varying values of $N_v$, the number of distillation modes. All tested cases below are with bilocal two-hadron operators in the rest frame. The $N\pi$ operator used is $p\,(\bar u\Gamma u-\bar d\Gamma d)+2\,n\,\bar d\Gamma u$, where $p=\epsilon_{abc}(u_a^\top C\gamma_5 d_b)u_c$, $n=\epsilon_{abc}(d_a^\top C\gamma_5 u_b)d_c$ and $\Gamma=\gamma_4\gamma_5$. The two dinucleon cases studied are isovector scalar operator $pp$ (${}^1S_0$, $I=1$) and isoscalar axialvector $(pn-np)/\sqrt2$ (${}^3S_1$, $I=0$, corresponding to a Deuteron operator), both in the upper Dirac components of the Dirac-Pauli basis. The three baryon case we study corresponds to the triton nucleus ($^3_1H$) $p\,[n^\top C\gamma_5\,n]$ with the proton spin open.}
\label{tab:wick_identities}
\begin{tabular}{lrrrr}
\toprule
 & & & \multicolumn{2}{c}{Classes}\\
System & Terms & $N_v$ & $u\neq d$ & $u=d$\\
\midrule
$N\pi$ & 120 & 10 & 120 & 19\\
$NN$, ${}^1S_0$ & 48 & 12 & 16 & 15\\
$NN$, ${}^3S_1$ & 144 & 12 & 36 & 18\\
$NNN$, $p\,[nn]$ & 2880 & 18 & 728 & 698\\
\bottomrule
\end{tabular}
\end{table}

The number of proportionality classes is the minimum number of Wick terms that has to be computed to reconstruct the full correlator faithfully. For the three-nucleon correlator most of the reduction in minimal number of Wick terms to be evaluated comes from the kernel and spin-matrix symmetries alone $2880\to728$, whereas isospin symmetry imposition contributes only to a marginal reduction from $728\to698$. The computational overhead in this step is determined by the choice of $N_v$, the number of distillation modes used to realize the primitive tensors. Therefore a small $N_v$ value could be used to reduce the overhead, while ensuring that the proportionality classes identified at a chosen $N_v$ also holds for larger values. A rule of thumb for a safe choice is $N_v\ge m=2\times$ (number of perambulators in a term), which we observe to hold at every larger value of $N_v$. Below this bound the test can lead to erroneous numerical grouping: at $N_v=4$ the three-nucleon terms with $u=d$ forms 299 classes, which are identities of smaller mode space that need not hold at higher mode space (\textit{e.g.} $N_v\ge18$, where there are 698 classes).

A random test cannot prove an identity, so \mypk\thinspace does not use it to change a plan. Instead, the optional compile pass \texttt{compile.wick\_aggregation}, which is off by default, merges two complete Wick terms only after it has found and checked an exact map between their tensors and indices, including the relative sign; the coefficients of merged terms are added exactly, and the proofs are stored in the plan and checked again whenever the plan is loaded. Flavors are treated as numerically equal only when this is declared with \texttt{numerical\_flavor\_groups}, only after the flavor-respecting Wick contractions are built, and only if the declared flavors read the same perambulator files. For the four systems of Table~\ref{tab:wick_identities}, the pass reaches exactly the class counts of the random test, and for the 18-operator $\pi\pi$ basis at $N_v=64$, merging with $u=d$ declared reduces the scheduled operations by 87\%. All other results in this paper are obtained without this pass.

\begin{table*}[bthp!]
\centering
\caption{A non-exhaustive list of \mypk\thinspace \emph{compile} and \emph{run} time options and their default values. }
\label{tab:options}
\begin{tabular}{lll}
\toprule
Option & Default & Meaning \\
\midrule
\multicolumn{3}{l}{\emph{Compile}}\\
\texttt{compile.spin\_basis} & \texttt{DR} & DeGrand-Rossi (\texttt{DR}) or Dirac-Pauli (\texttt{DP}) gamma matrices \\
\texttt{compile.scalar\_type} & \texttt{complex128} & double or single precision \\
\texttt{compile.contraction.strategy} & \texttt{greedy} & contraction-path search strategy  (Sec.~\ref{sec:execution}) \\
\texttt{compile.contraction.memory\_limit} & none & largest intermediate allowed in a contraction path \\
\texttt{compile.scheduling.strategy} & \texttt{memory\_aware} & order of the contraction steps \\
\texttt{compile.physical.reuse\_strategy} & \texttt{best\_fit} & reuse of memory buffers between steps \\
\texttt{compile.physical.memory\_limit\_bytes} & none & ceiling for the buffer memory of a work unit \\
\texttt{compile.wick\_aggregation.enabled} & \texttt{false} & exact merging of equal Wick terms (Sec.~\ref{sec:execution}) \\
\texttt{compile.numerical\_flavor\_groups} & none & flavors that read the same perambulators, e.g.\ $u$ and $d$ \\
\midrule
\multicolumn{3}{l}{\emph{Run}}\\
\texttt{run.device} & \texttt{cpu} & \texttt{cpu}, a CUDA device, or \texttt{auto} \\
\texttt{run.time\_batch\_size} & \texttt{auto} & number of work units evaluated together \\
\texttt{run.workers}, \texttt{run.threads\_per\_worker} & \texttt{auto} & worker processes and threads on a CPU node \\
\texttt{output.on\_existing} & \texttt{resume} & resume, overwrite, or stop at an existing output file \\
\midrule
\multicolumn{3}{l}{\emph{Optional}}\\
\texttt{data.kernel\_cache} & off & spatial kernels kept on disk for later runs \\
generalized perambulators & off & three-point functions (Sec.~\ref{sec:distillation_lowering}) \\
custom eigenvector provider & off & eigenvectors from user code instead of \texttt{HDF5} files \\
\texttt{concord inspect}, \texttt{--hardware} & -- & cost report of a plan; check against the machine \\
\texttt{concord run --run-record}, \texttt{--report} & -- & timing record and report of a run \\
\bottomrule
\end{tabular}
\end{table*}

Although the order in which the tensors of a connected graph are contracted does not change the value of the graph, it determines the FLOP count, the size and the reuse of the intermediate components, and the cost of computation. \mypk\thinspace offers six search strategies (\texttt{greedy}, \texttt{dynamic-programming}, \texttt{auto-hq}, \texttt{reuse-aware}, \texttt{adaptive}, and \texttt{auto}), that are identified with the \emph{compile} stage option \texttt{compile.contraction.strategy}, to identify efficient tensor contraction paths. The default path search strategy, \texttt{greedy} uses \texttt{opt\_einsum}'s fast, deterministic local-cost search. The strategies \texttt{dynamic-programming} and \texttt{auto-hq} use more involved search algorithms within \texttt{opt\_einsum} and thus demand higher compilation times, which is a one time overhead. For networks with more than 16 tensors, \texttt{auto-hq} is a randomized search, the best of 128 randomized \texttt{greedy} runs, so two compilations can give somewhat different plans; a run always uses its saved plan. The \texttt{reuse-aware} strategy compares four candidate plans for the whole correlation matrix, one from each search (\texttt{greedy}, \texttt{dynamic-programming} and two branch searches of \texttt{opt\_einsum}), each after partial-intermediate reuse, scheduling and memory planning, and saves the one with the fewest FLOPs whose buffer memory stays within a ceiling, by default that of the \texttt{greedy} plan. The compute time resource requirements grow exponentially with the number of tensors for search algorithms using \texttt{dynamic-programming} strategy. Thus \texttt{reuse-aware} falls back to the \texttt{greedy}-based plan for the whole correlation matrix when any of its networks has more than twelve tensors, which is the case for two-nucleon systems, whose networks have 22 tensors. The \texttt{adaptive} strategy assesses the \texttt{greedy}-based and \texttt{reuse-aware}-based plan performances on the actual hardware with synthetic tensors and engages the faster plan at run time. The \texttt{auto} strategy lets \texttt{opt\_einsum} choose the search for each network according to its size: an exhaustive search for up to four tensors, branch searches for up to 14, and \texttt{greedy} for larger networks. 

\mypk\thinspace plans memory use before reading any data. The basic job, a work unit, is one gauge configuration with one set of source and sink timeslices. At compile time, the contraction steps are put in an order that frees inputs early, and each result is assigned to a reusable buffer, so the memory a work unit needs is fixed in the plan and known in advance. The remaining choices are made at run time, after a safety reserve is subtracted from the available memory. Several work units can be batched together for efficient use of resources, either with user-specified batch sizes or with batch size chosen automatically (\texttt{run.time\_batch\_size: auto}, the default). On a CPU node, the pending work units are divided among worker processes that share the loaded plan; by default (\texttt{run.workers: auto}) the number of processes and of threads per process is chosen from the operations and memory of a work unit, the available cores and the free memory. \mypk\thinspace assumes the default hardware to be CPU node unless specified. The run time option \texttt{run.device: auto} identifies the availability of GPU resources and executes the job in the GPU if the plan fits there safely. 

Below, we discuss the example for a triton correlator, which has 2880 Wick contraction graphs, of which 2528 are fully connected nine-quark graphs. Of the remaining 352 graphs, 96 are those in which the proton propagates independently of the two-neutron system, and 272 are those in which one of the neutrons propagates independently of the remaining proton--neutron system. Sixteen graphs belong to both groups and correspond to all three nucleons propagating independently of each other. The hard part is not in constructing these graphs, but in choosing an \emph{order} in which to carry out the many contractions, since a poor choice can result in huge computing and memory requirements. 

We recommend compiling plans with several values of \texttt{compile.contraction.strategy} and examining each prepared plan with the optional call \texttt{concord inspect}. This optional call returns a report on the FLOP and memory requirement for each plan, which aids in assessing the cost of each plan and thus in choosing the plan that optimally utilizes the available resources. For the above example, at $N_v=32$, the default \texttt{greedy} plan for the triton correlator requires about 3~TB RAM, whereas the \texttt{auto-hq} plan requires only about 30~GB and about a hundred times fewer operations per work unit, suggesting the applicability of the latter plan. 

When executing each work unit, \mypk\thinspace first reads the perambulators from their \texttt{HDF5} files, converting them to the canonical order and conventions described in Sec.~\ref{sec:distillation_lowering}, and builds the spatial kernels from the supplied distillation modes, together with the gauge links when derivative or displaced operators are involved. The planned operations are executed in fixed order, using \texttt{PyTorch}~\cite{Paszke:2019xhz} on a CPU or a single GPU. All correlators of a job are written to one \texttt{HDF5} file. A work unit is marked complete only after all its correlators are written, so an interrupted job resumes where it stopped. A run record is given out that provides differential time spent in loading and checking the plan and making the run-time choices, data preparation, contraction, input waits and output. Table~\ref{tab:options} lists some of the \emph{compile} and \emph{run} time options along with their default values, where applicable. The \emph{compile} time option values are fixed in the plan, whereas \emph{run} time option values for a given plan are flexible across different runs and different hardware.

\section{Numerical validation}
\label{sec:validation}

A correlation matrix evaluated can be erroneous or corrupted in ways no single element reveals. A wrong relative sign between Wick terms, a transposed spin matrix, a complex-conjugated phase or a mismatched storage axis can all still yield smooth, plausible-looking correlators. Any additional layer of validation would hence be very valuable. We therefore verify the correlators produced by the compiled plans in two independent ways:
\begin{enumerate}
\item \emph{Using independent contraction code}: \mypk\thinspace produced correlators or operator kernels are checked against those produced by independently realized contraction codes utilizing other libraries and/or software suites such as \texttt{Chroma}~\cite{Edwards:2004sx}, \texttt{chroma\_laph} \cite{chromalaph}, and \texttt{NumPy}~\cite{Harris:2020xlr}. 
\item \emph{Identities and correlator symmetries}: Equivalence and similarities between correlators for different channels, such as those that might arise in the isospin limit, etc., can be used to verify the expected correlator symmetries. 
\end{enumerate}
Except for the hardware comparison at the end of this section, we use gauge configurations with $24^3$ spatial sites, one source timeslice and a limited number of sink timeslices, and at most 64 distillation modes. The tests on three-point correlator data use synthetic small grid lattice data. Table~\ref{tab:validation} summarizes the channels investigated based on  contraction topology, which can be extended to other channels with similar topology. For correlation functions/matrices of basic meson and baryon operators, bilocal two-meson operators, the derivative and gauge-transported meson operators and local diquark-antidiquark operators, the comparisons are made with results evaluated using contraction codes available/written in the \texttt{Chroma} software suite or the \texttt{chroma\_laph} package. Correlation functions of baryon-meson systems, the \(\Delta^{++}\) and those involving self-annihilating disconnected quark loops are validated using independent \texttt{NumPy} contraction realization. Two-nucleon correlators and the triton are checked with independent \texttt{NumPy} contraction codes and, for two nucleons, also through exact identities based on flavor equivalence. We discuss these comparisons further below. 

\begin{table*}[t]
\centering
\caption{Validation for different Wick contraction topologies presented in Fig.~\ref{fig:topologies} between \mypk\thinspace and independent evaluations. All comparisons were made with \texttt{complex128} arithmetic and agree up to double precision, with relative deviations of about $10^{-14}$ or smaller; the one exception, local mesons at rest, is explained in the text.}
\label{tab:validation}
\footnotesize
\setlength{\tabcolsep}{4pt}
\begin{tabular}{@{}ll@{}}
\toprule
Topology & Remarks \\
\midrule
single meson (a) & $\pi^+$, $\rho^+$ at rest and in moving frames up to \(\vec P^2=4\), all polarizations \\
 & Derivative and displaced bilinears $\bar d\gamma_5 \nabla_{+i} u$, $\bar d\gamma_5 T_{+i}u$, $\bar d\gamma_5 T_{+x}T_{+y}u$ \\
quark loops (b) & \(\bar uu\) two-point function; scalar and \(\gamma_4\) loops at rest, scalar loop in motion \\
single baryon (c, d) & proton, proton with a gauge-transported quark; \(\Delta^{++}\) \\
 & \(\Delta^+ = \Delta^{++}\) and \(\Lambda = p\) for equal \(u, d, s\) perambulators; \(\Delta^+\), \(\Lambda\) \\
two mesons (e, f) & bilocal two-meson operators, 13 channels at rest and in motion \\
 & local four-quark operators, 13 channels (up to the operator normalization) \\
baryon--meson (g) & \(p\pi^+\), \(N\pi\) (\(I_3=+\tfrac12\)), \(N\sigma\), \(N\pi\pi\), with equal-time propagation \\
two baryons & \(\tfrac{1}{\sqrt2}(np+pn) = pp\) for equal \(u, d\) perambulators; \\
& Trivially zero operator and correlator for \(I=0\) \({}^1S_0\); \\
 & \(NN\) \({}^1S_0\) and \({}^3S_1\), \(\vec p=0\) and \(\pm\vec p\); triton \(p\,[n\,C\gamma_5 n]\) \\
three-point & flavor-changing point current, five spin structures, synthetic data \\
\bottomrule
\end{tabular}
\end{table*}
The single-meson comparisons cover the rest frame, moving frames up to $\vec P^2=4$, the diagonal spin polarizations, and the spin polarization cross-correlations. These polarization tests check that the gamma-matrix and sign conventions are used consistently. All elements agree to double precision, except those that contain a local meson at rest, alone or inside a bilocal operator, which agree to $3\times10^{-9}$ and $7\times10^{-9}$, respectively. With its built-in eigenvector reader, \mypk\thinspace assumes that the stored eigenvectors are orthonormal and replaces the local zero-momentum overlap $V^\dagger V$ by the identity, which saves work; the results then depend on how well the stored eigenvectors are orthonormal. \texttt{Chroma}-based code evaluates $V^\dagger V$ from the stored eigenvectors, which in this data set are orthonormal only approximately \cite{Padmanath:2018tuc,Padmanath:2022cvl}. When $V^\dagger V$ is evaluated from the stored eigenvectors also in \mypk, these elements agree to $6\times10^{-15}$ and $1\times10^{-14}$. We have also validated a number of meson operators involving derivatives or displacements with the respective realization in the \texttt{chroma\_laph} package. Correlators of local four-quark operators with a diquark-antidiquark color structure also agree with contraction codes written in the \texttt{Chroma} software suite \cite{Prelovsek:2025vbr}, up to an overall constant that comes from the normalization of the operators: \mypk\thinspace projects every operator onto definite momentum with the plain sum $\sum_{\mathbf x}e^{+i\mathbf p\cdot\mathbf x}$, without a volume factor, as in Eq.~\eqref{eq:bilinear_spatial_kernel}. The baryon-meson correlators, including those with equal-time quark propagation (such as those in Fig.~\ref{fig:topologies}g), are also validated using independently realized contraction codes \cite{Lang:2016hnn}. 

For the $\Delta^+$ and the $\Lambda$ we check, in addition to comparisons with independent \texttt{NumPy} contraction codes, exact identities that hold when the quark masses are equal. With the diquark structures
\begin{align}
D(q_1,q_2)&=\epsilon_{abc}\,(q_{1a}^TC\gamma_5\,q_{2b}), \qquad \nonumber \\
D_i(q_1,q_2)&=\epsilon_{abc}\,(q_{1a}^TC\gamma_i\,q_{2b}),
\end{align}
and the operators
\begin{align}
\Delta^+_i &= \tfrac{1}{\sqrt3}\left[2D_i(u,d)\,u + D_i(u,u)\,d\right],\nonumber\\
\Lambda &= \tfrac{1}{\sqrt6}\left[2D(u,d)\,s + D(u,s)\,d - D(d,s)\,u\right],\nonumber
\end{align}
the relations
\begin{equation}
\langle \Delta^+_i \bar\Delta^+_j\rangle = \langle \Delta^{++}_i\bar\Delta^{++}_j\rangle,
\qquad
\langle \Lambda\bar\Lambda\rangle = \langle p\,\bar p\rangle
\label{eq:baryon_identities}
\end{equation}
hold exactly when the mass of $u$, $d$ and $s$ quarks are the same and thus their propagations are represented by the same perambulator. This is a nontrivial test on the relations between the compared correlators, as they are built from very different numbers of Wick contraction graphs: 8 versus 6 for the $\Delta^{+}$ versus $\Delta^{++}$, and 9 versus 2 for $\Lambda$ versus the proton. 

For correlations involving bilocal two nucleons, we validate all isospin and $S$-wave spin-parity sectors, including symmetry forbidden isoscalar scalar dinucleon operator that is trivially zero, based on results from independent contraction codes and additionally ensure two exact identities. The correlator of the isovector scalar dinucleon operator has 144 Wick contraction graphs, and in the isosymmetric limit it should be equal to the $pp$ correlator, which has 48 Wick terms. This equality is exact and must be satisfied for each configuration. The isoscalar scalar dinucleon operator corresponds to the equivalent antisymmetric combination, which vanishes at the operator level itself, for any quark masses, so that its Wick terms must cancel exactly. Both these identities are validated to a tolerance of $\sim10^{-16}$, essentially the limit of computer precision, on two configurations. In addition, the $I=1$ $^1S_0$ and $I=0$ $^3S_1$ correlation matrices of two operators, with both nucleons at rest and with back-to-back momenta, agree with an independent \texttt{NumPy} contraction code to $10^{-15}$, and the triton correlator $p\,[n\,C\gamma_5\,n]$, with 2880 Wick terms, agrees to $10^{-14}$.

We also validate \mypk's three point correlation evaluation results using propagator data generated on a synthetic $4^3\times4$ lattice with three colors. With $N_v=6$ random orthonormal vectors per timeslice in place of Laplacian eigenvectors, we built ordinary perambulators for all source times and the generalized perambulators of Eq.~\eqref{eq:generalized_peram} for $\Gamma\in\{\gamma_1,\gamma_2,\gamma_3,\gamma_4,1\}$. For four different orderings of the sink, current and source times and all five currents, the three-point functions agree with an independent evaluation with the full propagators to $10^{-14}$. This check validates the Wick contraction and its sign, the orientation of the generalized perambulator, the elementals and their phases, and the binding of user-supplied generalized data. The calculation of generalized perambulators lies outside the package.

To check the portability, the complete $64\times64$ source-sink grid for a rest-frame pion and a positive-parity nucleon was computed using the same plans on a 64-core CPU node and on a single NVIDIA H100. The results from both jobs agree to $5.4\times10^{-15}$ (pion) and $1.7\times10^{-15}$ (nucleon), relative to the channel magnitude. The data validation reported here is limited to what was actually compared. The following features are also available in the \mypk\thinspace suite but have not been tested and are not listed in Table~\ref{tab:validation}:
\begin{itemize}
    \item disconnected loops other than the scalar and $\gamma_4$ loops of Table~\ref{tab:validation};
    \item derivative and transported operators with other spin structures, longer paths or higher derivatives, and baryons with other displaced quarks;
    \item baryon-meson and baryon-two-meson channels with large $N_v$ values;
    \item execution on GPUs for correlators other than the pion and the nucleon.
\end{itemize}

\section{Performance Evaluation}
\label{sec:performance}

\begin{figure*}
\centering
\includegraphics[width=\textwidth]{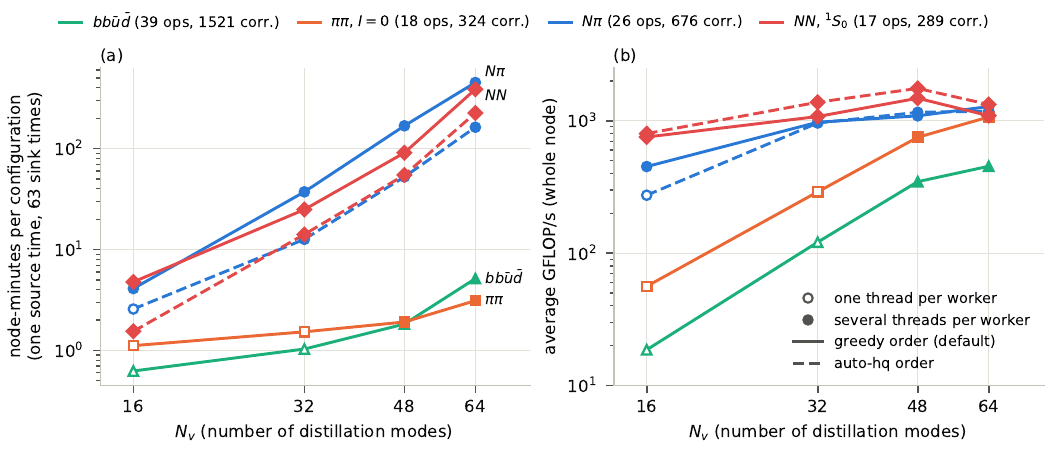}
\caption{Wall time and average arithmetic rate in the evaluation of correlation matrices for channels listed in Table~\ref{tab:benchmark_bases} for varying number of distillation modes. All measurements are based on runs on a single 64-core CPU node, with the default settings (in Table~\ref{tab:options}) and automatically chosen worker processes. (a) Node minutes for one configuration and one source time with 63 sink times. (b) Average arithmetic rate of the whole node in giga floating-point operations per second (GFLOP/s). Solid lines: \texttt{greedy} contraction paths (the default); dashed lines: \texttt{auto-hq} contraction paths. Open symbols: one thread per worker; filled symbols: several threads per worker.}
\label{fig:throughput}
\end{figure*}

The cost of a distillation calculation is dominated by the contractions, and the practical question for a user is how this cost grows when the operator basis grows, how much cost advantage would the \mypk\thinspace package give. We answer these questions for a set of two-hadron systems and for triton nucleus, for which the correlation matrices contain various contraction topologies of Fig.~\ref{fig:topologies}. The estimates in this section are based on the analytical cost report of the compiled plan, prepared by the \texttt{concord inspect} call. The measured run times on a CPU node are given at the end of this section (Fig.~\ref{fig:throughput}).

To make the performance evaluation, we utilize one $N_f=3$ gauge  configuration with geometry $24^3\times64$. The distillation subspace is spanned by up to 64 eigenvectors of the discretized gauge-covariant 3D Laplacian. The light and heavy quark perambulators are supplied for all 64 source times externally. Flavor symmetry only affects the proportionality classes across various allowed Wick contractions, and the timings do not depend on the physics of these choices.

All two-hadron operators studied are bilocal in nature with definite momentum assignments for both position locality groups in the operator. The three-hadron operators are also accordingly tri-local. The isodoublet baryon channel studied includes local proton operators as well as six displaced-proton operators along with 19 bilocal $N\pi$ operators. We list all the channels studied and the corresponding operator bases utilized in Table~\ref{tab:benchmark_bases}. The $^4$He operator is listed only for its number of quark pairings. 

\begin{table*}[t]
\centering
\caption{Channels considered for performance benchmarking. $N_{\rm op}$ is the size of the full basis; ``Wick terms'' counts the effective number of terms of the full correlation matrix after exact cancellation.
$\pi^0=(\bar u\gamma_5u-\bar d\gamma_5d)/\sqrt2$, $\sigma=(\bar uu+\bar dd)/\sqrt2$,
$N_{C\gamma_5}=\epsilon_{abc}(u_a^TC\gamma_5d_b)q_c$.}
\label{tab:benchmark_bases}
\footnotesize
\begin{tabular}{@{}lp{0.52\textwidth}rrl@{}}
\toprule
Channels & Operators & \(N_{\rm op}\) & Wick terms & Topologies \\
\midrule
$N$, \(N\pi\) \((I=\tfrac12)\) & \(p\); \(p\) with transported third quark (6); \(\sqrt{2/3}\,n(\vec p)\pi^+(-\vec p)+\sqrt{1/3}\,p(\vec p)\pi^0(-\vec p)\), \(|\vec p|^2\le2\) (19) & 26 & 47\,142 & exchange, equal-time \\
\(\pi\pi\), \(I=0\) & \(\sigma\); \((\pi^+\pi^-+\pi^-\pi^++\pi^0\pi^0)/\sqrt3\) at \(\pm\vec p\), \(|\vec p|^2\le4\) (17) & 18 & 60\,410 & loops, vacuum \\
\(bb\bar u\bar d\), \(J^P=1^+\) & \(\epsilon\epsilon\,(b^TC\gamma_3b)(\bar u C\gamma_5\bar d^T)\); \(B B^*(\pm\vec p)\) and \(B^*B^*(\pm\vec p)\), \(|\vec p|^2\le2\) (19 each) & 39 & 11\,858 & local four-quark \\
\(NN\), \(^1S_0\), \(I=1\) & \(p(\vec p)\,C\gamma_5\,p(-\vec p)\), \(|\vec p|^2\le4\) & 17 & 13\,872 & six-quark exchange \\
\(NN\), \(^3S_1\), \(I=0\) & \([p(\vec p)\,C\gamma_3\,n(-\vec p)-n(\vec p)\,C\gamma_3\,p(-\vec p)]/\sqrt2\), \(|\vec p|^2\le4\) & 17 & 41\,616 & six-quark exchange \\
\(^3\)H & \(p(x)\,[n(y)\,C\gamma_5\,n(z)]\), \(\vec p=0\) & 1 & 2\,880 & nine-quark exchange \\
\(^4\)He & \([p(x)\,C\gamma_5\,p(y)]\,[n(z)\,C\gamma_5\,n(w)]\), \(\vec p=0\) & 1 & 518\,400 & twelve-quark exchange \\
\bottomrule
\end{tabular}
\end{table*}

Plans were built on an Intel Xeon Gold 6438N, except for triton nucleus, for which we used an Intel Xeon Platinum 8358. All measurements were made on a single CPU node with two 32-core Intel Xeon Platinum 8358 processors and 256~GB of memory. Software versions utilized in these tests were Python 3.13 and \texttt{PyTorch} 2.13, in combination with \texttt{opt\_einsum} 3.4 and \texttt{h5py} 3.16. All correlators were evaluated using \texttt{complex128} arithmetic, and for a single source timeslice and 24 to 63 sink timeslices $t\neq t_0$. Compute time consumed in the actual numerical evaluation was taken from the run record of \texttt{concord run}. The plan was compiled independently beforehand and the associated compile times are reported separately. We quote the node time needed for one configuration and source time.

\begin{figure*}[pos=!b]
\centering
\includegraphics[width=\textwidth]{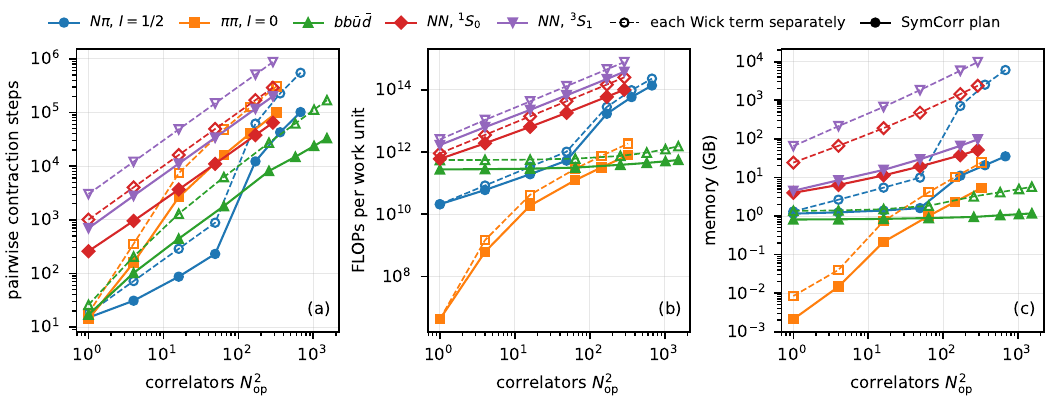}
\caption{Cost report of the compiled plans for the channels listed in Table~\ref{tab:reuse_full} as a function of the number of correlators $N_{\rm op}^2$. The estimates are for $N_v=64$ and \texttt{greedy} contraction paths. The order of operators is as listed in Table~\ref{tab:benchmark_bases}. (a) Number of pairwise contraction steps, (b) contraction FLOPs per work unit, (c) peak memory requirement. Dashed: (a, b) every distinct connected network contracted separately, without sharing partial intermediates; (c) every value of the compiled plan kept until the end. Solid: the compiled \mypk\thinspace plan: (a, b) after partial-intermediate reuse and (c) after scheduling and buffer reuse.}
\label{fig:reuse_scaling}
\end{figure*}

Figure~\ref{fig:reuse_scaling} shows how the cost of computation grows with the number of correlators $N_{\rm op}^2$ at a fixed $N_v=64$ for the first five channels listed in Table~\ref{tab:benchmark_bases}. In Table~\ref{tab:reuse_full}, we list the full compiled plan summary and various resource requirements. The dashed lines in Fig.~\ref{fig:reuse_scaling} denote estimates for the strategy, where every distinct network is contracted separately, without sharing partial intermediates between networks, and every value of the plan is kept in memory until the end. The solid curves represent the computation resource requirements while using \mypk's compiled plan. There are a few effects evident from these figures. 

First, the \mypk\thinspace plan provides a clear improvement in the computation requirements in comparison with this baseline: for the full bases it needs 1.7 to 2.8 times fewer FLOPs (Table~\ref{tab:reuse_full}). This reduction is prominent in bases, where there are bilocal interpolators. The most expensive components in different channels are generally different, owing to the different natures of Wick terms involved in each sector. For channels, where cost is dominated by one expensive operator, adding cheap operators is almost free. In the doubly bottom tetraquark ($bb\bar u\bar d$) basis, the local diquark-antidiquark operator kernel comes with four distillation mode indices. Its diagonal correlator alone needs $2.8\times10^{11}$ FLOPs for $N_v=64$ (scaling as $N_v^6$ with \texttt{greedy} contraction path strategy). However, a meson-meson correlator grows as $N_v^3$.

The memory requirement depends on the channel being studied. In the two-baryon systems the memory requirement reduces by more than an order of magnitude with the \mypk\thinspace compiled plan. The reduction factor is further observed to grow with the basis size. With the full basis considered, for the isovector scalar dinucleon ($NN$ $I=1$ $^1S_0$), this factor is 46, whereas the same for isoscalar vector dinucleon ($NN$ $I=0$ $^3S_1$, alias the Deuteron) is 101. A cost reduction is also observed in $\pi\pi$ and $bb\bar u\bar d$, although it is marginal, with a maximum factor of around five with full basis considered. A na\"ive storing of every tensor involved in the \texttt{greedy}-plan for the isodoublet nucleon channel with the full basis including $N\pi$ operators would consume 6.1~TB per work unit, whereas the default memory-aware scheduling and buffer reuse bring the requirement of the same plan down to 35~GB.

\begin{table*}[t]
\centering
\caption{Composition and cost of correlation matrices in five different channels. The estimates are provided for the case $N_v=64$ and assuming \texttt{greedy} contraction paths. The pairs of numbers in the networks and steps columns are the unique and the total numbers of tensor networks and of pairwise contraction steps; a step contracts two tensors into one (Sec.~\ref{sec:execution}). Estimates for FLOPs per work unit are listed for the case when distinct networks are contracted separately $f_i$ as well as for the case that exploits partial-intermediate tensor reuse $f_f$ in the form $f_i\to f_f~(f_i/f_f)$. The memory requirement for the plan is listed considering all tensors stored $m_i$ and only the scheduled physical peak $m_f$, with the reduction factor in parentheses in the form $m_i\to m_f~(m_i/m_f)$. The rightmost column lists the compile time on one core, with up to 32 compilations running in parallel.}
\label{tab:reuse_full}
\footnotesize
\setlength{\tabcolsep}{3pt}
\begin{tabular}{@{}lrrrrrllr@{}}
\toprule
Channels & \(N_{\rm op}\) & corr. & Wick terms & networks & steps & FLOPs & memory (GB) & compile (min)\\
\midrule
$N\pi$, $I=1/2$ & 26 & 676 & 47\,142 & 36457/70588 & 101869/550973 & \(2.29\times10^{14}\)\,\(\to\)\,\(1.38\times10^{14}\) (1.7) & 6085\,\(\to\)\,35.1 (173) & 85 \\
$\pi\pi$, $I=0$ & 18 & 324 & 60\,410 & 25716/142062 & 99672/308260 & \(1.86\times10^{12}\)\,\(\to\)\,\(7.92\times10^{11}\) (2.4) & 24\,\(\to\)\,5.3 (5) & 132 \\
$bb\bar u\bar d$ & 39 & 1521 & 11\,858 & 17634/17634 & 33095/169322 & \(1.54\times10^{12}\)\,\(\to\)\,\(5.56\times10^{11}\) (2.8) & 6\,\(\to\)\,1.2 (5) & 4 \\
$NN$, $^1S_0$ & 17 & 289 & 13\,872 & 13872/13872 & 64576/291312 & \(2.52\times10^{14}\)\,\(\to\)\,\(1.01\times10^{14}\) (2.5) & 2379\,\(\to\)\,51.3 (46) & 6 \\
$NN$, $^3S_1$ & 17 & 289 & 41\,616 & 41616/41616 & 200794/873936 & \(7.84\times10^{14}\)\,\(\to\)\,\(3.81\times10^{14}\) (2.1) & 9809\,\(\to\)\,96.9 (101) & 54 \\
\bottomrule
\end{tabular}
\end{table*}

Table~\ref{tab:strategies} presents the cost report of the contraction paths suggested by various search strategies on one or more representative bases per channel. Plans are assessed taking into account the memory ceiling of the CPU node (256 GB). For the single nucleon operator with $N_v=64$ listed in the first row, the \texttt{greedy} path forms a larger set of intermediate objects and demands 4.8 times more FLOP and 8 times more memory than the plans formed by other strategies. With a larger basis for the channel, involving single proton local and displaced-quark operators along with bilocal $N\pi$ operators, the improvement with plan formed with non-\texttt{greedy} strategies suggest reduction in the FLOP requirement by a factor of around 9.8 and a reduction in memory requirement by a factor 8. For the two-nucleon correlators, there is a modest improvement in the FLOP and memory requirement with non-\texttt{greedy} strategies with a maximum improvement by a factor of 3. We observe that the plan built with any of the strategies for the $N\pi$ as well as $NN$ channels are feasible on a CPU node. In the case of di-meson ($\pi\pi$) and tetraquark ($bb\bar u\bar d$) systems, we observe that different strategies form plans with only modest differences. All formed plans are feasible on a CPU node. 

For the triton nucleus ($^3H$) correlations, not all the strategies lead to feasible plans. With $N_v=32$ the \texttt{greedy} strategy, the plan requires $2.65\times10^{15}$ FLOP to be performed and a peak RAM requirement of about 2.9~TB, which is substantially more than the available RAM. The \texttt{auto-hq} strategy suggests plan that needs $2.78\times10^{13}$ FLOP and only 30~GB RAM requirement per work unit, which fits well within a single CPU node. The RAM requirement with plan from \texttt{auto-hq} exceeds the memory ceiling with $N_v=64$ and thus cannot be executed with our present resources. Notably, the one-time expense of building a plan, which is the compile time, consumed by the \texttt{auto-hq} strategy is around 600~s, about eight times that of the \texttt{greedy} strategy. Thus the substantial run time improvement in the triton correlation function evaluation clearly trades off the larger compile time resource consumption between \texttt{auto-hq} and \texttt{greedy} strategies. It is also evident that for the triton, \texttt{reuse-aware} strategy falls back to the \texttt{greedy} search resulting in the plan requiring 2860~GB peak RAM at $N_v=32$, which is not feasible on the node, and thus we had to resort to the plan suggested by the \texttt{auto-hq} search strategy.

\begin{table*}[t]
\centering
\caption{\texttt{concord inspect} report for plans suggested by various contraction-path search strategies at fixed basis size. For a given channel at a fixed basis size, there are three numbers presented for each strategy (\texttt{greedy}, \texttt{dynamic-programming}, \texttt{auto-hq}, \texttt{reuse-aware}) in the form $a/b/c$, where $a$ refers to the scheduled FLOPs per work unit, $b$ refers to physical peak memory in GB and $c$ refers to the \emph{compile} time in seconds. The \texttt{reuse-aware} entry is the plan with the fewest FLOPs among the \texttt{greedy}, \texttt{dynamic-programming} and two branch-search plans whose memory stays within that of the \texttt{greedy} plan, or the \texttt{greedy} plan when a network has more than twelve tensors (Sec.~\ref{sec:execution}); here it coincides with the \texttt{greedy} or the \texttt{dynamic-programming} entry.}
\label{tab:strategies}
\footnotesize
\setlength{\tabcolsep}{3pt}
\begin{tabular}{@{}lrrllll@{}}
\toprule
Channels & \(N_{\rm op}\) & \(N_v\) & \texttt{greedy} & \texttt{dynamic-programming} & \texttt{auto-hq} & \texttt{reuse-aware} \\
\midrule
$N\pi$ & 1 & 64 & \(2.09\times10^{10}\) / 1.15 / 0 & \(4.33\times10^{9}\) / 0.145 / 0 & \(4.33\times10^{9}\) / 0.145 / 0 & \(4.33\times10^{9}\) / 0.145 / 0 \\
$N\pi$ & 7 & 64 & \(5.30\times10^{11}\) / 1.6 / 1 & \(5.43\times10^{10}\) / 0.197 / 1 & \(5.43\times10^{10}\) / 0.197 / 1 & \(5.43\times10^{10}\) / 0.197 / 2 \\
$\pi\pi$ ($I=0$) & 8 & 64 & \(1.25\times10^{11}\) / 0.999 / 172 & \(1.65\times10^{11}\) / 0.28 / 283 & \(1.65\times10^{11}\) / 0.28 / 310 & \(1.25\times10^{11}\) / 0.999 / 298 \\
$bb\bar u\bar d$ & 8 & 64 & \(3.16\times10^{11}\) / 0.883 / 4 & \(2.56\times10^{11}\) / 4.86 / 10 & \(2.56\times10^{11}\) / 4.86 / 10 & \(3.16\times10^{11}\) / 0.883 / 6 \\
$NN$ ($^1S_0$) & 4 & 64 & \(6.46\times10^{12}\) / 11.3 / 9 & \(4.59\times10^{12}\) / 4.32 / 404 & \(4.24\times10^{12}\) / 4.28 / 94 & \(6.46\times10^{12}\) / 11.3 / 14 \\
$NN$ ($^3S_1$) & 2 & 64 & \(6.31\times10^{12}\) / 8.45 / 7 & \(2.92\times10^{12}\) / 3.23 / 223 & \(2.53\times10^{12}\) / 3.56 / 70 & \(6.31\times10^{12}\) / 8.45 / 11 \\
$^3$H & 1 & 16 & \(2.05\times10^{13}\) / 117 / 74 & -- & \(4.70\times10^{11}\) / 2.25 / 611 & -- \\
$^3$H & 1 & 32 & \(2.65\times10^{15}\) / 2860 / 74 & -- & \(2.78\times10^{13}\) / 30.5 / 602 & \(2.65\times10^{15}\) / 2860 / 119 \\
$^3$H & 1 & 48 & -- & -- & \(2.66\times10^{14}\) / 127 / 612 & -- \\
$^3$H & 1 & 64 & \(1.67\times10^{17}\) / 80\,200 / 75 & -- & \(1.49\times10^{15}\) / 407 / 604 & -- \\
\bottomrule
\end{tabular}
\end{table*}

Figure~\ref{fig:throughput} shows the measured cost of computing correlations for the $N\pi$, $\pi\pi$, $bb\bar u\bar d$ and $NN$ (${}^1S_0$) bases of Table~\ref{tab:benchmark_bases} on one CPU node for varying $N_v=16$ to 64. All settings are the default values listed in Table~\ref{tab:options}, except that the \emph{run} time option \texttt{run.time\_batch\_size} is set to the value 1, a single timeslice per batch. The contraction paths are those of the \texttt{greedy} strategy, and for the baryon systems also of \texttt{auto-hq} (dashed lines), with the worker processes and their threads chosen automatically. The quantity shown is the node minutes for one configuration and one source time with 63 sink times, taken from the wall time excluding the overhead time spent in loading the plan (the most expensive cases were run with 24 or 32 sink times and are scaled to 63). 

The four systems fall into two groups. The $\pi\pi$ and $bb\bar u\bar d$ plans consist of many small contractions, so their time is set by the number of contraction steps rather than by the arithmetic. A configuration takes 0.6 to 5 node-minutes, and the time grows only by factors of two and five from $N_v=32$ to 64. Such plans run best as many single-thread workers, \textit{e.g.} for $bb\bar u\bar d$ at $N_v\le32$ the automatic choice is 63 workers with one thread each, which gives 5 to 12 times the throughput of the same plan in a single process. For $\pi\pi$ at $N_v\le32$, and for some $N\pi$ plans, \mypk\thinspace cannot tell from its rough memory estimate that all workers fit on the node; it then first lets each worker evaluate one work unit while measuring its memory, and splits the remaining units afterwards. In this benchmark, with one configuration and one source time, the second round makes the $\pi\pi$ jobs 1.5 to 1.7 times longer than a single round of 63 workers; in production, with many configurations per job, it is paid only once. The $NN$ and $N\pi$ plans are dominated by large contractions. Their time grows roughly as $N_v^{3.6}$ to $N_v^{4}$, from four to five node-minutes at $N_v=16$ to about seven node-hours at $N_v=64$, and since each work unit needs more memory, fewer workers with more threads are started as $N_v$ grows: 16 workers with four threads up to $N_v=32$, and three to five workers at $N_v=64$. For systems involving baryons, we observe the choice of the contraction path also pays at run time. With \texttt{auto-hq} strategy (dashed lines), the \emph{compile} time overhead is $\sim$3-5 times more compared to that with the \texttt{greedy}-strategy, whereas the execution wall time for the same correlation matrices with \texttt{auto-hq}-plan is approximately 1.6 to 3.2 times less than that required by the \texttt{greedy}-plan on the same platform.

Panel (b) presents the average arithmetic rate of the whole node in GFLOP per second (GFLOP/s): four real floating-point operations for each FLOP of the cost report, which counts a complex multiply--add as two, divided by the node time of panel (a). It generally rises with $N_v$ and reaches 1 to 1.8~TFLOP/s for the cases involving baryons with $N_v\ge32$. For comparison, the nominal double-precision peak of this node, two 32-core Intel Xeon Platinum 8358 processors with two AVX-512 FMA units per core at the base frequency of 2.6~GHz, is about 5.3~TFLOP/s. The plans made of small contractions reach only 0.02 to 0.3~TFLOP/s at $N_v\le32$, where the fixed cost of each contraction step, and for $bb\bar u\bar d$ also the writing of its 1521 correlators, limits the speed; this is where we expect the largest gains from further development. These measurements describe the present version of \mypk\thinspace on one node and are meant to show the trend with $N_v$. Execution on a GPU has been verified (Sec.~\ref{sec:validation}), but its performance has not been studied.

\section{About the software}
\label{sec:software}

\mypk\thinspace 1.0 is a pure Python package, released under the MIT License, compatible with Python version 3.11 or higher and is available at \url{https://github.com/Tanishk-Shrimal/ConCorD}~\cite{ConCorD}. Its run-time dependencies are \texttt{NumPy}~\cite{Harris:2020xlr}, \texttt{PyTorch}~\cite{Paszke:2019xhz} (CPU or CUDA builds), \texttt{h5py}~\cite{Collette:2013}, \texttt{opt\_einsum}~\cite{Smith:2018opteinsum}, \texttt{Pydantic}, \texttt{PyYAML} and \texttt{Typer}. There are no additional requirements such as compiled extension, MPI library or vendor-specific GPU code. Therefore, the package and the compiled plan are portable across a wide range of platforms covering personal laptops, office desktops,  CPU cluster nodes or CUDA GPU nodes. In case of multiple available hardware resources, the package allows the choice of the device at the run time. The independent \texttt{NumPy} contraction codes of Sec.~\ref{sec:validation} and the scripts that compare them with \mypk\thinspace are provided in \texttt{tools/}, together with a script that writes random input data on which they can be repeated. The validation results are summarized in the user guide, \texttt{docs/user\_guide.md}. The lattice data used here are described in the data availability statement at the end of the paper.

The user is expected to provide a \texttt{job.yaml} file, which passes on the operator factory and its parameters, the lattice extents, the compile options ($N_v$, scalar type, spin basis, contraction strategy), the run options (configurations, source and sink times, device, time batch size) and the data sources (eigenvectors, smeared gauge links, perambulators and their storage layout) to \mypk. The workflow proceeds in two stages, with an optional inspection stage in between. Given the \texttt{job.yaml}, \texttt{concord plan} prepares and builds a portable \texttt{plan.concord} file, which holds the compiled plan together with a fingerprint of the operator source (including its local Python helpers), the factory parameters and the compile options; data paths, configurations, time contexts and the device are excluded, so one plan can be compiled once and run on many configurations and machines.
\begin{verbatim}
concord plan job.yaml      # compile -> plan.concord
\end{verbatim}

The plan saved in \texttt{plan.concord} is portable across different hardware platforms. The plan cost and its feasibility on a given machine can also be inspected, before the real evaluation using the command
\begin{verbatim}
concord inspect plan.concord --hardware
\end{verbatim}
This call provides the analytical cost report discussed in Sec.~\ref{sec:performance} that includes various details covering number of Wick terms to be evaluated, unique tensor networks and intermediate tensor components, FLOP and memory requirement, etc. With the \texttt{--hardware} option, it also checks whether the memory the plan needs fits the CPU and GPU memory of the machine.

The second stage is the numerical execution, which follows the syntax below and produces an output \texttt{HDF5} file containing the requested correlation matrix. 
\begin{verbatim}
concord run job.yaml       # execute -> results.h5
\end{verbatim}
\texttt{concord run} rejects a plan whose fingerprint no longer matches the job and resumes an interrupted output file.

The architecture of \mypk\thinspace and the ideas behind its features are the authors' own. They were implemented with AI coding assistants: Codex (OpenAI) in the initial development and, later, Claude Code with Claude Opus~5.5 and Claude Fable~5.2 (Anthropic), which were also used for the contraction codes of the cross-checks in Sec.~\ref{sec:validation}. We verified the code with physics-motivated tests, comparing correlators with those of independent codes and checking exact identities between correlators (Sec.~\ref{sec:validation}), and with logical tests of its components. We take full responsibility for the published code.

\section{Limitations and Future Development}
\label{sec:limitations}
The scope of validated measurements with \mypk\thinspace 1.0 was presented in Table~\ref{tab:validation}. Correlators involving disconnected and self-annihilating quark loops with more complex spin structures, nontrivial momentum assignments or displaced quark fields have not yet been validated. Baryon operators such as those defined and constructed in Refs.~\cite{Basak:2005ir,Edwards:2011jj} have not been tested on \mypk\thinspace 1.0, which one may explore if excited baryon structures are being studied. Temporal derivatives within the operator definitions are not supported and thus rejected explicitly. Operator bases involving glueball interpolators are also not supported.

The \emph{compile} time in search for an exact optimal contraction path using the \texttt{dynamic-programming} strategy grows exponentially with the number of tensors involved. The \texttt{reuse-aware} strategy therefore compares its four searches only when every network of the correlation matrix has at most twelve tensors, and otherwise uses the \texttt{greedy} plan. Equalities that rely on symmetries in the correlation matrices (for example $C_{ij}=C_{ji}^{\ast}$) are not exploited, although equalities from kernel and spin-matrix symmetries are exploited. Operators whose quarks are all at one point, such as a local six-quark operator, lead to rank-six kernels in the distillation mode space resulting in large memory requirements. For such operators, we plan to avoid building the kernel in the distillation mode space. Instead, the operator is evaluated at a sample of spatial positions, as in the position-space sampling of Refs.~\cite{Stump:2024lqx,Stump:2025owq}, and the Wick contractions at each sampled position are computed together as determinants~\cite{Detmold:2012eu,Doi:2012xd,PhysRevD.110.114505}. Contraction paths are chosen for each network separately, without regard to the intermediates it could share with other networks, except in the \texttt{reuse-aware} strategy for small networks. The memory known in advance is that of the buffers of the plan; the workspace used while the kernels are built is not included. Execution on a GPU has been verified against the CPU results for the pion and the nucleon (Sec.~\ref{sec:validation}), but a systematic performance study on GPUs was not possible with the hardware available and is left for future work. 

Natural extensions of \mypk\thinspace are to lift the present limitations listed above. The extensions that fit the present single-CPU or single-GPU mode of execution include noise-diluted estimators such as stochastic Laplacian Heaviside \cite{Morningstar:2011ka}, stochastic estimation of disconnected contributions with dedicated variance-reduction techniques \cite{Bali:2009hu}; optimized distillation profiles \cite{Knechtli:2022bji}; reuse-driven contraction-path selection across the correlation matrix \cite{10.1145/3592979.3593409}. Multi-GPU and multi-node execution of a single work unit and out-of-core host storage are not implemented and would support more complex hadronic systems and larger lattice grid sizes. Most of these extensions change only one or two of the layers numbered in Fig.~\ref{fig:workflow}, because each layer passes a fixed representation to the next. For example, optimized distillation profiles weight the distillation indices of the kernels, so they change only how the kernels are defined and built (layers 2 and 6), while the operator language, the Wick contractions and the planning stay the same. Likewise, the exact merging of Wick terms (Sec.~\ref{sec:execution}) is one optional pass in layer 3, and a new contraction-path search changes only layer 4.

\section{Summary}
\label{sec:summary}

We have presented \mypk, a compiler package that turns a symbolic quark-level description of hadron interpolating operators into an optimized evaluation of distillation correlation matrices. All steps between the operator definitions and the numbers are automatic: Wick contraction with fermionic signs, lowering to distillation tensor networks, reuse across the whole correlation matrix, contraction-path selection, scheduling, memory planning and execution on CPUs or GPUs. The compiled plan is independent of the numerical data and of the hardware. The framework supports local, displaced and gauge-transported quark fields, single- and multi-hadron operators, disconnected loops, and three-point functions with point currents through user-supplied generalized perambulators.

A number of basic correlators built using \mypk\thinspace were validated with independently realized contraction codes. Correlators for mesons, two-meson operators, a local four-quark operator and derivative and gauge-transported mesons in the rest as well as moving frames agree up to the machine precision, modulo cases where \mypk\thinspace uses identity matrix for $V^\dagger V$, whereas the deviation from identity matrix on this quantity is determined by the precision with which the eigenvectors in $V$ were determined. The correlations were also verified for baryon-meson systems, $\Delta$ and $\Lambda$ baryons, two-nucleon systems, and the triton nucleus. Correlators with self-annihilating quark loops at the source and sink timeslices and a three-point correlation function with a point current also agree with independent contractions. We also verify exact flavor identities configuration by configuration between differently contracted correlators, namely $\langle\Delta^+_i\bar\Delta^+_j\rangle=\langle\Delta^{++}_i\bar\Delta^{++}_j\rangle$, $\langle\Lambda\bar\Lambda\rangle=\langle p\,\bar p\rangle$, and, for two nucleons, the equality of the $I=1$, $I_3=0$ correlator with $pp$, together with the vanishing of the Pauli-forbidden $I=0$ spin-singlet combination. 

The package also offers several strategies for finding an efficient plan. We demonstrate the application of these strategies across a number of topical channels and discuss the feasibility of studying them on a CPU node. The main limitations of the package are that contraction paths are chosen per network, that the memory model does not yet include kernel-construction workspaces, and that a single work unit runs on one device (Sec.~\ref{sec:limitations}).

\section{Acknowledgments}

We are grateful to our colleague, Bhabani Sankar Tripathy, for various discussions. We also thank John Bulava and Colin Morningstar for making available the \texttt{chroma\_laph} package \cite{chromalaph}, which was utilized in generating the distillation dataset necessary for the development of \mypk. We have utilized the \texttt{Chroma} software suite \cite{Edwards:2004sx}, the \texttt{chroma\_laph} package \cite{chromalaph}, and the \texttt{NumPy}~\cite{Harris:2020xlr} library to realize independent contraction codes involved in crosschecks. We also acknowledge the use of the Python libraries \texttt{PyTorch}~\cite{Paszke:2019xhz}, \texttt{opt\_einsum}~\cite{Smith:2018opteinsum} and \texttt{h5py}~\cite{Collette:2013}. We acknowledge the use of the HPC facility at IMSc Chennai. M.P. gratefully acknowledges support from the Department of Science and Technology, India, ANRF(SERB earlier) Start-up Research Grant No. SRG/2023/001235 and the Department of Atomic Energy, India.

\section*{Declaration of competing interest}

The authors declare that they have no known competing financial interests or personal relationships that could have appeared to influence the work reported in this paper.

\section*{Data availability}

The source code of \mypk, the independent \texttt{NumPy} contraction codes used for the validation and a summary of the validation are available in the \mypk\thinspace repository (Sec.~\ref{sec:software}). Additionally, the repository comes with a script that generates random input data of any size, on which the comparisons with the \texttt{NumPy} codes can be repeated and the computation time estimated: they depend only on the sizes of the data, not on their values. The lattice data used in this work (Laplacian eigenvectors and perambulators) were generated by the authors; they are available on reasonable request.

\section*{Declaration of generative AI and AI-assisted technologies in the manuscript preparation process}

During the preparation of this work the authors used Claude (Anthropic) in order to draft and revise parts of the text, improve its language, and write the code for the plots, diagrams and flowcharts. After using this tool, the authors reviewed and edited the content as needed and take full responsibility for the content of the publication.

\bibliographystyle{unsrt}
\bibliography{references}

\end{document}